\documentclass{emulateapj}
\usepackage{graphicx}
\submitted{}
\journalinfo{Submitted to the Astrophysical Journal}
\defcitealias{cohen11}{CC11}

\shorttitle{$Chandra$ View of Abell 2443}
\shortauthors{Clarke, Randall, Sarazin, Blanton \& Giacintucci}

\newcommand{\gsim}{\mathrel{\hbox{\rlap{\lower.55ex \hbox{$\sim$}} \kern-.3em \raise.4ex \hbox{$>$}}}}

\begin{document}

\title{$Chandra$ View of the Ultra-Steep Spectrum Radio Source in
  Abell 2443: Merger Shock-Induced Compression of Fossil Radio Plasma?}

\author{T.~E.~Clarke\altaffilmark{1}, S.~W.~Randall\altaffilmark{2},
  C.\ L.\ Sarazin\altaffilmark{3}, E.~L.~Blanton\altaffilmark{4},
  S. Giacintucci\altaffilmark{5,6}} \altaffiltext{1}{Naval Research
  Laboratory, Code 7213, Washington, DC, 20375 USA,
  tracy.clarke.ca@nrl.navy.mil} \altaffiltext{2}{Harvard Smithsonian
  Center for Astrophysics, 60 Garden Street, Cambridge, MA 02138, USA}
\altaffiltext{3}{Department of Astronomy, University of Virginia, PO
  Box 400325, Charlottesville, VA 22904-4325}
\altaffiltext{4}{Astronomy Department and Institute for Astrophysical Research, Boston
  University, Boston, MA 02215, USA} \altaffiltext{5}{Department of
  Astronomy, University of Maryland, College Park, MD 20742-2421 USA}
\altaffiltext{6}{Joint Space-Science Institute, University of
  Maryland, College Park, MD 20742-2421 USA}

\begin{abstract}
We present a new $Chandra$ X-ray observation of the intracluster
medium in the galaxy cluster Abell 2443, hosting an ultra-steep
spectrum radio source. The data reveal that the intracluster medium is
highly disturbed. The thermal gas in the core is elongated along a
northwest to southeast axis and there is a cool tail to the north. We
also detect two X-ray surface brightness edges near the cluster
core. The edges appear to be consistent with an inner cold front to
the northeast of the core and an outer shock front to the southeast of
the core. The southeastern edge is coincident with the location of the
radio relic as expected for shock (re)acceleration or adiabatic
compression of fossil relativistic electrons.

\end{abstract}

\keywords{ galaxies: clusters: general -- galaxies: clusters:
  individual (Abell 2443) -- radio continuum: galaxies -- X-rays: galaxies: clusters}

\section{Introduction}

Structure formation is thought to occur in a hierarchical manner with
objects forming through the gravitational collapse of initial density
enhancements and subsequent growth through accretion and merging. The
largest gravitationally bound objects in the Universe, galaxy
clusters, are thought to evolve through multiple mergers, which occur
at the intersection of large-scale structure filaments. A typical
major cluster merger involves collisions at velocities $\ga$ 2000 km
s$^{-1}$ and can release a total kinetic energy 10$^{63-64}$ ergs
\citep{sarazin00}. These mergers are the most energetic events in the
Universe since the Big Bang and can drive strong shocks into the
intracluster medium (ICM). To date, shocks have been clearly
identified in only a small number of merging systems (e.g., Bullet,
\citealt{markevitch02}; Abell 520, \citealt{markevitch05}; Abell 2146,
\citealt{russell10}; Abell 3667, \citealt{fino10}; Abell 754,
\citealt{macario11}; Abell 2744, \citealt{owers11}; MACS0744.8+3927,
\citealt{korgut11}; RXCJ1314.4-2515, \citealt{mazzotta11}; Abell 521,
\citealt{bourdin13}). The shock energy is dissipated through
compression of magnetic fields, particle acceleration or
reacceleration, and heating of the ICM. In the presence of magnetic
fields, the accelerated relativistic particles can be observable in
the radio regime as diffuse synchrotron emission, although the
efficiency of shock acceleration is very uncertain
\citep[e.g.][]{kang12}.

A growing number of clusters of galaxies are known to host regions of
extended radio synchrotron emission which have no optical
counterparts. These radio sources have low surface brightness, large
sizes ($\gsim$ 500 kpc) and steep spectra \citep[see review
  by][]{ferrari08}. They have been observationally classified broadly
as halos and relics. Giant radio halos (called radio halos hereafter)
are located at cluster centers, show rather regular structure
and little or no polarized emission. Relics are seen in
projection near the cluster peripheries; they are generally elongated
in shape and are often highly polarized. Roughly 80 clusters have been
confirmed to contain halos and or relics \citep{feretti2012}, but the origin and evolution
of this diffuse emission is still a matter of debate. Both types of
diffuse emission are seen only in clusters undergoing mergers. A
third class of diffuse radio emission, called mini-halos, is seen only
in more relaxed, cool core clusters.

Radio halos are thought to be the result of either collisions between
cosmic-ray protons and thermal protons \citep{dennison80} or particle
acceleration through turbulence injected into the merging cluster
system \citep[see review by][]{petrosian08}. 

Models for the formation of radio relics rely on the presence of a
shock, driven by a cluster merger, within the X-ray gas. The two
primary mechanisms for forming radio relics are (i) diffusive shock
acceleration (DSA) through the Fermi-I mechanism and (ii) adiabatic
compression of fossil radio plasma by the merger shock. In the first
scenario the relativistic electrons are the result of either
acceleration of thermal electrons \citep{ensslin98} or reacceleration
of pre-existing cosmic ray particles \citep{kang11} which have
experienced multiple crossings of the shock front to gain energy. In
the adiabatic compression scenario, the merger shock is thought to
pass over and compress a cocoon of aged relativistic plasma where the
high sound velocity inside the old radio plasma would inhibit the
shock \citep{ensslin01}. Compression of the cocoon would increase the
magnetic field strength and energy density of the relativistic
electrons within the cocoon. This would lead to a boost in the
synchrotron luminosity of the emission during this `Flashing' phase of
the model of \citet{ensslin01}. Such a re-ignited cocoon would thus
become observable if it is sufficiently young that the cutoff
frequency from losses remains above the observing
frequency. \citet{ensslin01} estimate that for re-ignition, the fossil
plasma should not be older than 0.2 Gyr near the cluster center or up
to 2.0 Gyr on the cluster outskirts prior to the shock passage.

The acceleration mechanism of relics may (in some cases) be
differentiated by the radio spectrum. Shock accelerated or
reaccelerated relics are expected to be fit to a power law spectrum
while adiabatically compressed relics should have a strongly curved
spectrum reflecting their pre-compression losses. \citet{keshet10} has
proposed a unified model for radio halo and relic formation based on a
single secondary cosmic ray electron model, where the time evolution
of both the magnetic fields and the cosmic ray distribution are taken
into account to explain the origin of the diffuse emission.

We have recently detected an ultra steep spectrum (USS) diffuse radio
source in Abell 2443 \citep[hereafter CC11]{cohen11}.  In this paper
we present results from a short $Chandra$ observation of the
system. This first pointed X-ray observation of this cluster reveals a
disturbed ICM, with two X-ray surface brightness edges. One of the
edges is consistent with a cold front, while the other edge is more
suggestive of a shock edge. The spatial coincidence of the latter edge
with the diffuse radio emission supports the interpretation of the
diffuse radio emission as a shock-induced (compressed or
reaccelerated) USS relic. We assume $H_o$ = 70 km $\rm{s}^{-1}$
Mpc$^{-1}$, $\Omega_M$ = 0.3, and $\Omega_\Lambda$ = 0.7 (1\arcsec\ =
1.95 kpc at $z = 0.108$) throughout. Errors are given at the 68\%
confidence level unless otherwise stated.

\section{The Galaxy Cluster Abell 2443}
\label{sect:A2443}

Abell 2443, a rich cluster at an intermediate redshift of $z = 0.108$
\citep{sr99}, and is dominated by two large galaxies near its
center \citep{crawford99}. The cluster is also host to the powerful
radio source 4C+17.89 which \citeauthor{crawford99} associate with the
fainter of the two central dominant galaxies.

\begin{figure}
\includegraphics[width=0.5\textwidth]{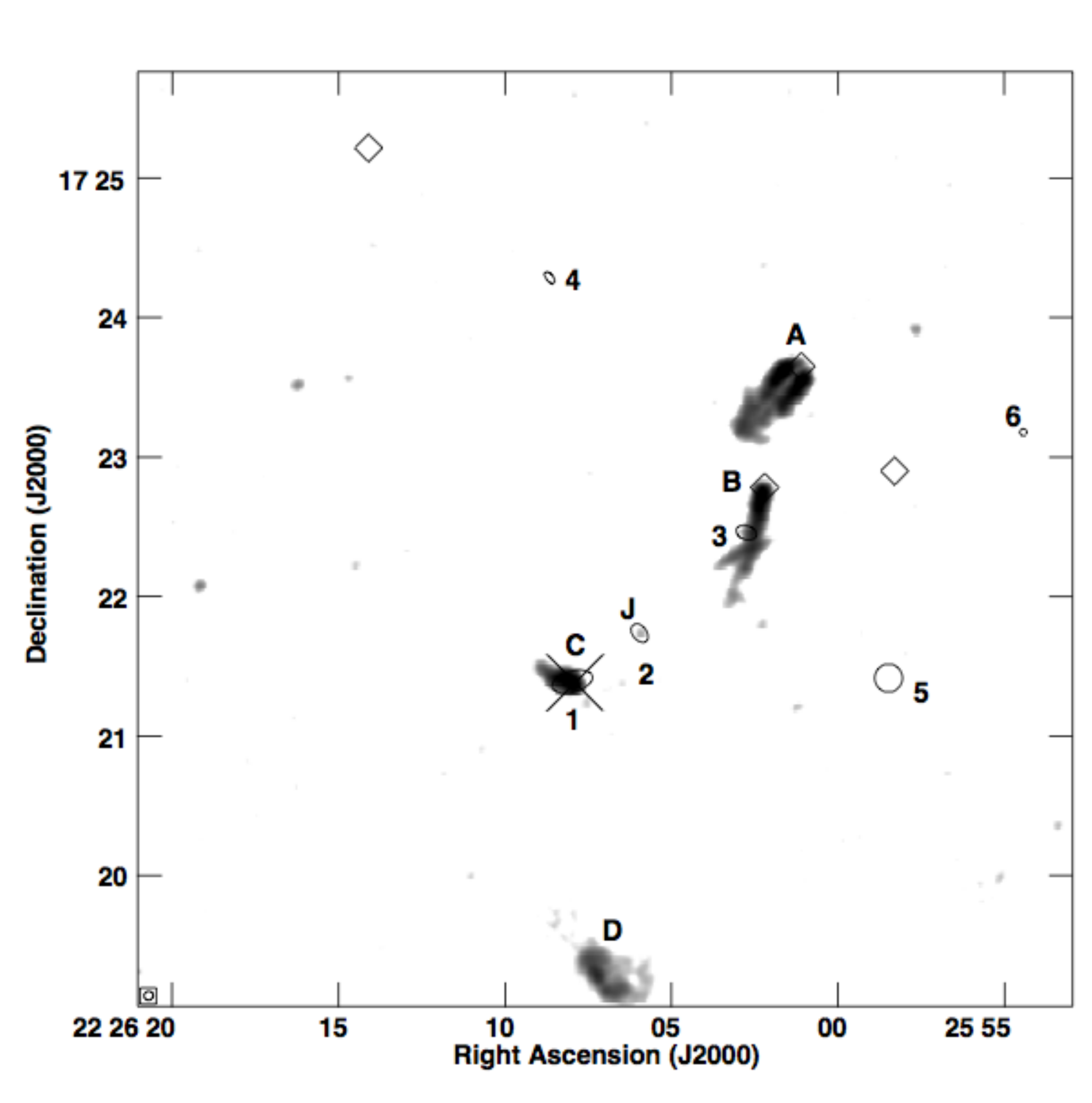}
\caption{\small {NRAO VLA 1425 MHz radio image in greyscale. The beam, with
  resolution of 4.00 $\times$ 3.75 arcsec, is shown in the lower left
  corner.  Radio source labels follow the convention of
  \citetalias{cohen11} and add new source J which is a faint radio
  source with an X-ray counterpart that is likely a background
  source. The six bright X-ray point sources discussed in
  \S~\ref{sect:compact} and Table~\ref{tbl:compact} are indicated by
  the corresponding numbered black ellipses. The BCG from
  \citet{crawford99} associated with source C is indicated by a
  cross. The diamonds indicate the four spectroscopically-confirmed
  cluster-member ellipticals that are within 1 magnitude in $I$-band
  of the BCG \citep{miller02}. Two of these cluster-member ellipticals
  are associated with the tailed radio galaxies A and B. The field of
  view is roughly 800 kpc on a side.
\label{fig:l_xray}}}
\end{figure}
Deep NRAO Very Large Array (VLA) radio observations of A2443 at 74,
325 and 1425 MHz reveal the presence of a diffuse USS radio source as
well as three tailed galaxies \citepalias{cohen11}. In
Figure~\ref{fig:l_xray}, we show the VLA 1425 MHz greyscale image of
the cluster with source labels. Two of the tailed radio galaxies (A \&
B in Figure~\ref{fig:l_xray}) are spectroscopically-confirmed cluster
members while the third source host (C in Figure~\ref{fig:l_xray}) has
a photometric redshift consistent with cluster membership and is
identified by \citet{crawford99} as the brightest cluster
galaxy. Source A displays classic narrow-angle-tail (NAT) morphology,
source B may be a head-tail source or an NAT that was not resolved by
\citetalias{cohen11}, while source C appears as a distinct core and
small tail to the northeast. NAT sources are generally located in the outer
regions of clusters which are dynamically complex
\citep{bliton98}. The bending of the tails is a result of either ram
pressure due to bulk motions from the merger event or the orbital
velocity of the host galaxy. Given the very similar tail directions
for sources A and B, it seems likely that the morphology is a result
of ram pressure due to a merger with a northwest to southeast component.

Supporting evidence of a disturbed dynamical state in Abell 2443 comes
from the multi-color photometry of \citet{wen07} where 301 cluster
member candidates (289 of them new candidates) are
identified. Candidates within 2 Mpc of Abell 2443 were found by
\citet{wen07} to be distributed along a northwest to southeast axis,
similar to the merger axis suggested by the radio tail orientation.

The USS source discovered in \citetalias{cohen11} has a spectral
index\footnote{Spectral index ($\alpha$) is defined such that
  $S_{\nu}\propto \nu^{\alpha}$ for a source with measured flux $S$ at
  frequency $\nu$.} between 74 MHz and 325 MHz of $\alpha_{74}^{325} =
-1.7\pm 0.1$ and much steeper spectral index between 325 MHz and 1425
MHz of $\alpha_{325}^{1425} = -2.8\pm 0.1$. The source has no clear
optical counterpart and is extended over a linear extent of $\sim$ 350
kpc if we assume it is at the redshift of the
cluster. \citetalias{cohen11} interpret the source as an USS radio
relic. USS relics are found only in merging clusters and are thought
to be connected to effects of merger shocks. Observations from the
ROSAT All-Sky Survey \citep{RASS} PSPC data revealed the presence of
X-ray emission associated with the cluster, but had insufficient
resolution to examine the cluster merger state \citepalias[see
  Figure~10 of][]{cohen11}.

\section{$Chandra$ Observations}
\label{sect:chandra}
\begin{figure*}[t]
\includegraphics[width=0.475\textwidth]{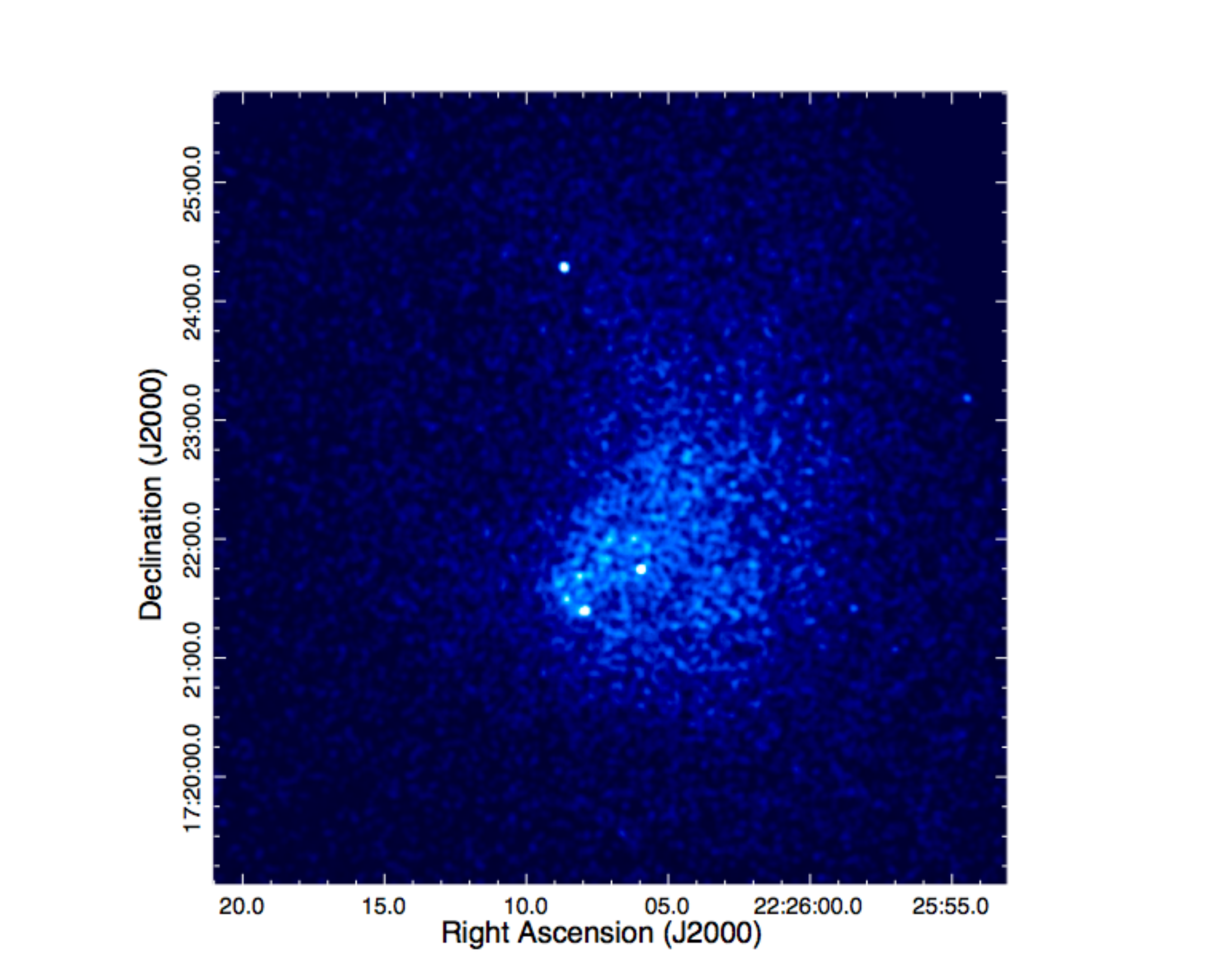} 
\hfill
\includegraphics[width=0.41\textwidth]{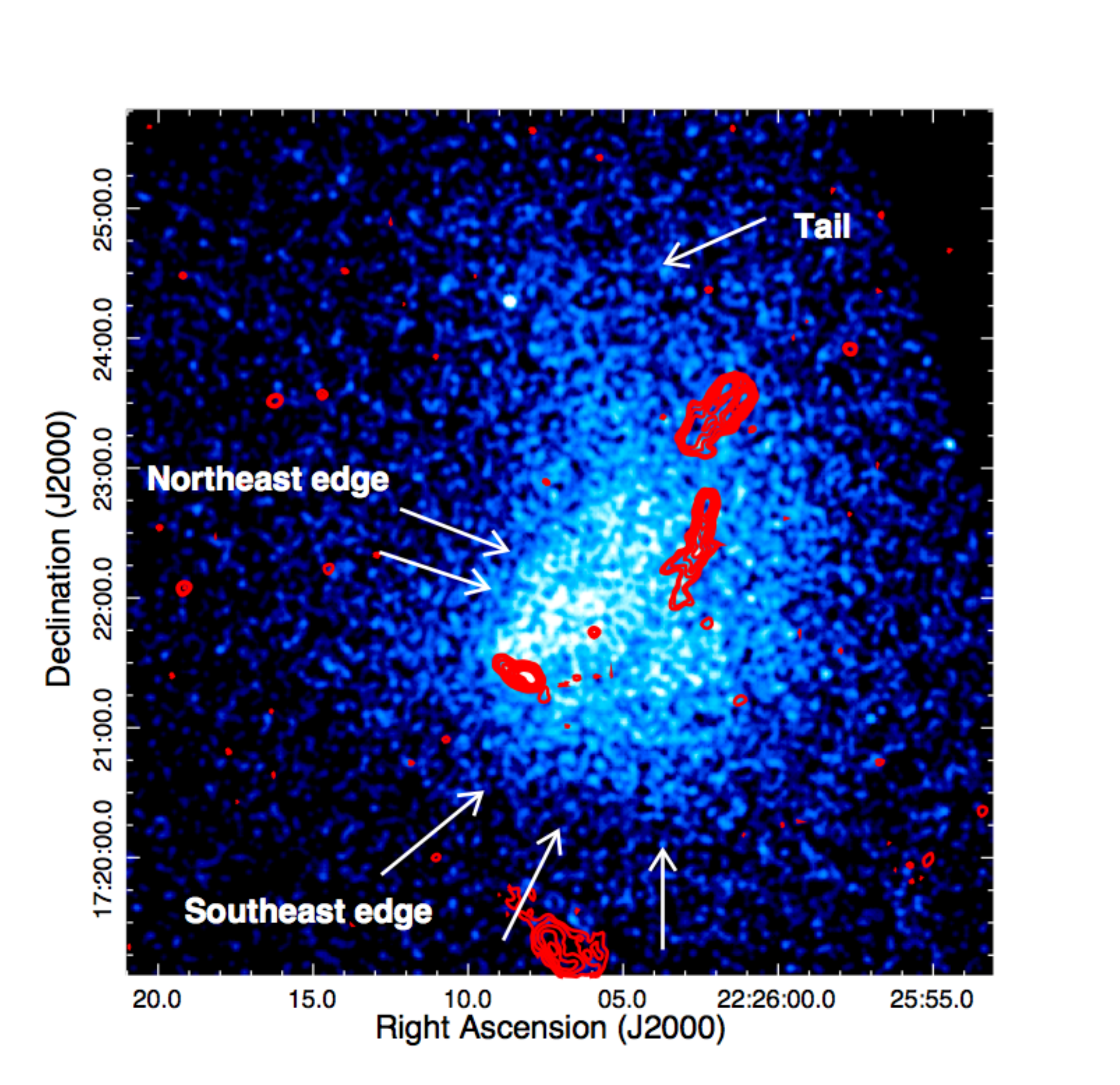} 
\caption{\small {Background corrected $0.3-7.0$ keV $Chandra$ ACIS-S images of
  Abell 2443 smoothed with a Gaussian of 1\farcs5. Left: The bright
  central ICM appears disturbed with an elongation along a southeast
  to northwest direction. Also visible are several compact X-ray
  sources including two near the center of the ICM. Right: Same as the
  left panel but with a logarithmic stretch to highlight the features
  of the diffuse emission. The northern tail as well as the northeast
  and southeast edges (all marked by arrows) are clearly visible. Overlaid are the
1425 MHz contours corresponding to the emission seen in Figure~\ref{fig:l_xray}. Both
  images show a region of 800 kpc on a side.
\label{fig:smo3}}}
\end{figure*}

We observed Abell~2443 with $Chandra$ for 15.2~ksec on 2010 August 20
(Obs ID 12257). The cluster core was placed on the
back-side-illuminated ACIS-S3 CCD and the data were taken in VFAINT
mode. All data were reprocessed from the level 1 event files using the
latest calibration files (as of {\sc CIAO 4.3}). CTI and
time-dependent gain corrections were applied. {\sc lc\_clean} was used
to remove background
flares\footnote{\url{http://asc.harvard.edu/contrib/maxim/acisbg/}}.
The energy range for the flare cleaning was $2.5-7$ keV as recommended
in the
COOKBOOK\footnote{\url{http://asc.harvard.edu/contrib/maxim/acisbg/COOKBOOK}}
and the filtering was done on the ACIS-S1 and ACIS-S3 CCDs with the
cluster contribution on the S3 CCD masked. The mean event rate was
calculated from a source free region using time bins within 3$\sigma$
of the overall mean, and bins outside a factor of 1.2 of this mean
were discarded. There were no periods of strong background flares. The
resulting cleaned exposure time was 14.5~ksec

Diffuse emission from Abell~2443 fills almost the entire S3 field of
view.  We therefore used the {\sc
  CALDB\footnote{\url{http://cxc.harvard.edu/caldb/}}} blank sky
background files appropriate for the observation, normalized such that
the count rate in the $10-12$~keV band matched the observed
rate. Normalizations were computed using data from the S3 and S1 chips
with point sources and regions of bright, diffuse cluster emission
excluded. Since bright emission from Abell 2443 fills most of the S3
chip, the normalization was largely determined by data from the S1
chip. To generate exposure maps, we assumed a MEKAL model with $kT =
5.0$~keV, Galactic absorption of $N_H=5.21 \times 10^{20}$ cm$^{-2}$
\citep{dl}, and an abundance of 30\% Solar at a redshift $z = 0.108$,
which is consistent with typical results for the extended emission
from detailed spectral fits (see \S~\ref{sect:spectral}).

\section{X-ray Imaging Analysis}
\label{sect:imaging}

\begin{figure}
\includegraphics[width=0.475\textwidth]{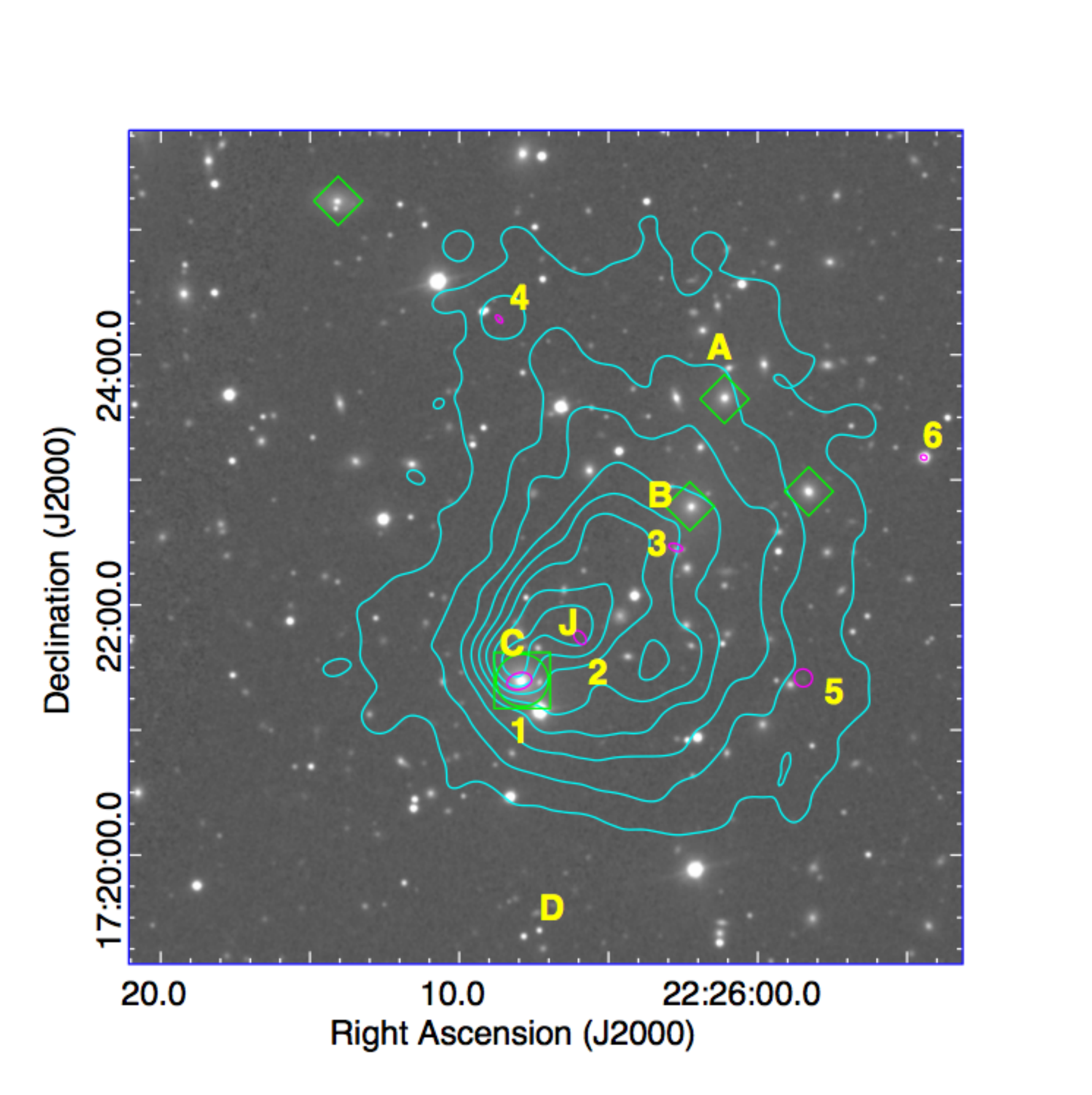}
\caption{\small {Sloan Digital Sky Survey (SDSS) $r$-band image of Abell 2443
  with smoothed $Chandra$ $0.3-7.0$ keV X-ray contours overlaid in
  cyan. Labels are as in Figure~\ref{fig:l_xray} except we have
  changed the BCG cross to a circle-box to avoid covering the optical
  emission.
\label{fig:opt_xray}}}
\end{figure}

The background corrected $0.3-7.0$ keV smoothed ACIS-S $Chandra$ image
of the central 800 kpc of Abell 2443 is shown in
Figure~\ref{fig:smo3}. The cluster is dominated by a large region of
diffuse emission with no obvious compact core. The main
characteristics seen for the diffuse emission are the edges to the
northeast and southeast of the core as well as a faint diffuse tail
extending to the north. Figure~\ref{fig:smo3} contains two panels with
different intensity scales to bring out the brighter emission in the
cluster core on a linear scale (left panel) and the fainter extended
emission using a logarithmic scale (right panel). The left panel
clearly shows the compact sources and central cluster emission
including an inner edge to the northeast while the right panel reveals
the details of the northern tail and the outer southeast surface
brightness edge. We label the main diffuse ICM features discussed in
this paper within the right panel of Figure~\ref{fig:smo3} and also
overlay the VLA 1425 MHz radio contours on this panel. In
Figure~\ref{fig:opt_xray} we show an overlay of the smoothed X-ray contours
on the SDSS $r$-band image of Abell 2443. We also include labels from 
Figure~\ref{fig:l_xray} for reference. 

\subsection{Compact Sources}
\label{sect:compact}

We searched for point sources on the ACIS-S3 chip using the wavelet
decomposition program {\sc wavdetect} \citep{freeman02}. We used
  a range of spatial scales covering 1, 2, 4, 8, and 16 pixels where
  the images were binned by a factor of two such that 1 pixel is
  roughly 1\arcsec. Sources were visually confirmed in the X-ray
images. We found a total of 6 sources (labeled in
Figures~\ref{fig:l_xray} \& \ref{fig:opt_xray}) and list details of
these sources in Table~\ref{tbl:compact}. Three of the brightest
sources are easily visible in Figure~\ref{fig:smo3}. Two bright
compact X-ray sources are near the cluster center and the third is
visible near the tail to the north. The eastern central point source
is co-incident with the cD galaxy PCG 068859 at a photometric redshift
of $z=0.107$ \citep{slee94}. This galaxy \citep[identified as the BCG
  by][]{crawford99} is host to the head-tail radio source C identified
by \citetalias{cohen11} (see Figure~\ref{fig:opt_xray}). The western
bright central X-ray point source is co-incident with a faint compact
1425 MHz radio source labeled source J in Figure~\ref{fig:l_xray} as
well as a faint optical galaxy seen in \citet{trujillo01}. Based on
the faint absolute magnitude ($I-$band magnitude of $-16.99$) it is
unlikely that this galaxy is an elliptical cluster member and thus it
is most likely an unrelated faint background object detected at all
three wavelengths. The SDSS photometric redshift of $z_{phot}=0.227\pm
0.091$ further supports the background source identification. The
bright northern X-ray source has no associated radio counterpart but
the optical counterpart has a photometric redshift of
$z_{phot}$=0.115$\pm$0.027 from the SDSS DR8, indicating that it is
likely a cluster member. Of the remaining three fainter X-ray sources,
one is co-incident with the tail of galaxy B in the notation of
\citetalias{cohen11}. This source is also co-incident with object 6 of
\citet{trujillo01} and is likely an elliptical cluster member based on
the measured magnitude, although we note that there are two optical
IDs in the SDSS within the X-ray source region and this is the lowest
significance {\sc wavdetect} source. The X-ray source to the west of
the cluster core has no radio counterpart and is identified in the
SDSS DR8 as a star. The final X-ray source to the NW of the cluster is
coincident, although slightly off-center, to a source in the USNO
UCAC3 catalog \citep{UCAC3}. The magnitudes listed in the catalog show
that the source is too bright optically to be a cluster member and the
colors are more consistent with an A-type star.

\begin{figure}
\includegraphics[width=0.5\textwidth]{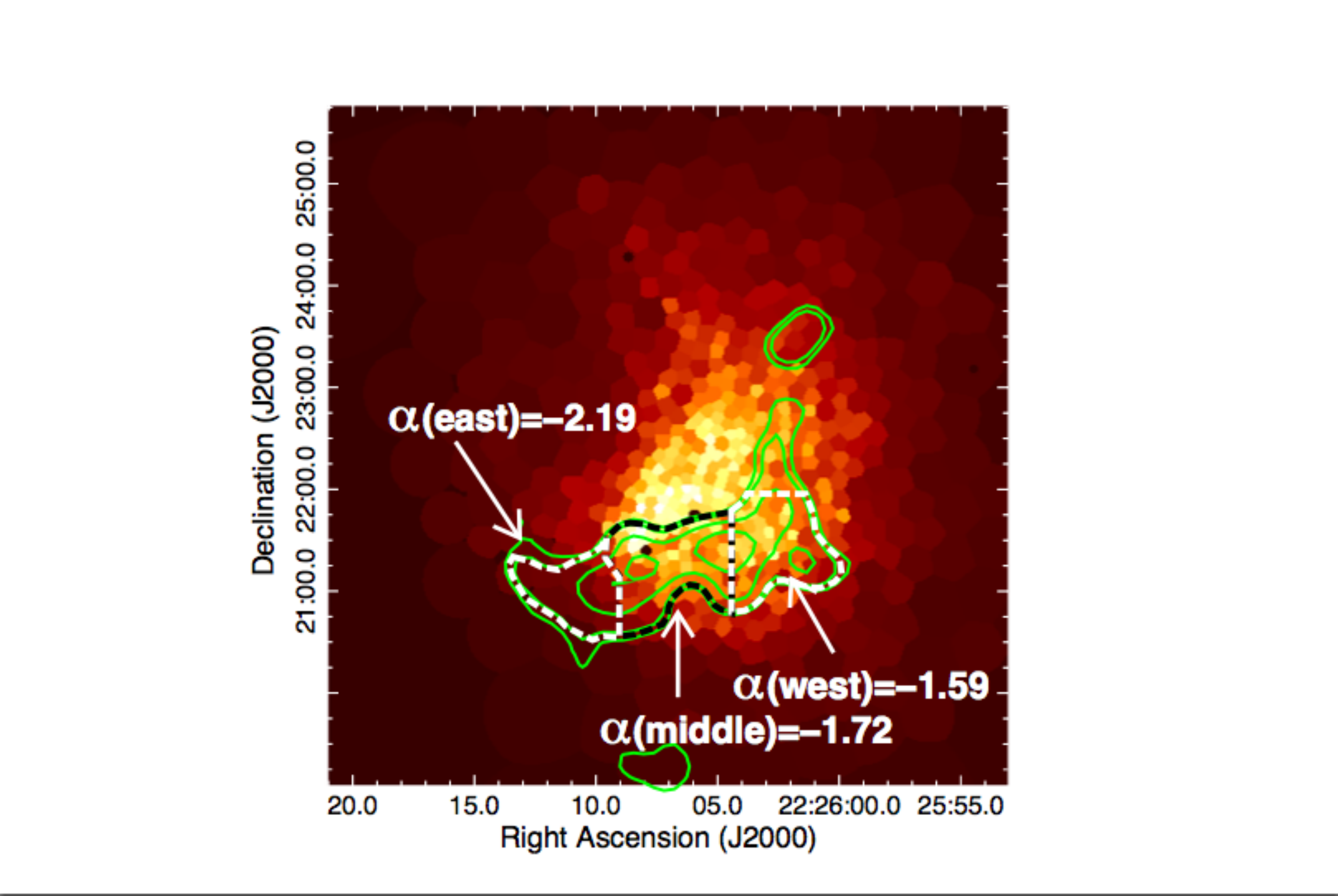}
\caption{\small {$Chandra$ $0.3-7.0$ keV image of the central region of Abell
  2443 adaptively binned using Voronoi tessellation with a S/N=5 to
  highlight the diffuse emission. Overlaid are VLA 74 MHz contours in
  green. The three contiguous dashed areas show the regions where the
  spectral index values discussed in \S~\ref{sect:radio} were
  calculated. The middle region (indicated in black) contains the
  compact source C. The spectral index between 330 MHz and 74
  MHz for each region is indicated.
\label{fig:vt_5_74}}}
\end{figure}

\subsection{Extended Emission}

In Figure~\ref{fig:vt_5_74} we show details of the diffuse
intracluster medium with the 74 MHz radio contours from
\citetalias{cohen11} overlaid. The X-ray image was made by
adaptively binning using the Voronoi tessellation technique of \citet{ds06}
with a signal-to-noise (S/N) of 5. The compact X-ray sources have been
masked in the process. The main X-ray features visible are the
tear-drop shaped region of diffuse X-ray emission with sharp edges to the
northeast and southeast of the core, and more gradual surface brightness decrease
to the west. The overall complex shape of the ICM emission and the
lack of a compact core are both consistent with a cluster undergoing a
merger. The diffuse intracluster X-ray emission extends over a region
of roughly 600 kpc $\times 550$ kpc and is elongated in the north-south
direction on large scales and in the northwest to southeast direction
within the central 150 kpc.

The sharp inner northeast edge has no associated radio emission while
the southeast edge is coincident with the location of the USS radio
emission discovered by \citetalias{cohen11}. The connection between
the USS radio emission and southeast edge suggests that this may be a
shock front which has compressed fossil radio plasma or
reaccelerated relativistic particles in this region. The
northern portion of the cluster is dominated by an X-ray tail that
arcs toward the northeast and may represent ram-pressure stripped
material from a merger, or the tail may be the outer
portion of the spiral excess from a sloshing core. We discuss these
possibilities further in Section~\ref{sect:discussion}.

\section{X-ray Spectral Analysis}
\label{sect:spectral}

\subsection{Average Cluster Spectrum}
\label{sect:cluster_spec}
\begin{figure*}[t]
\includegraphics[width=0.35\textwidth]{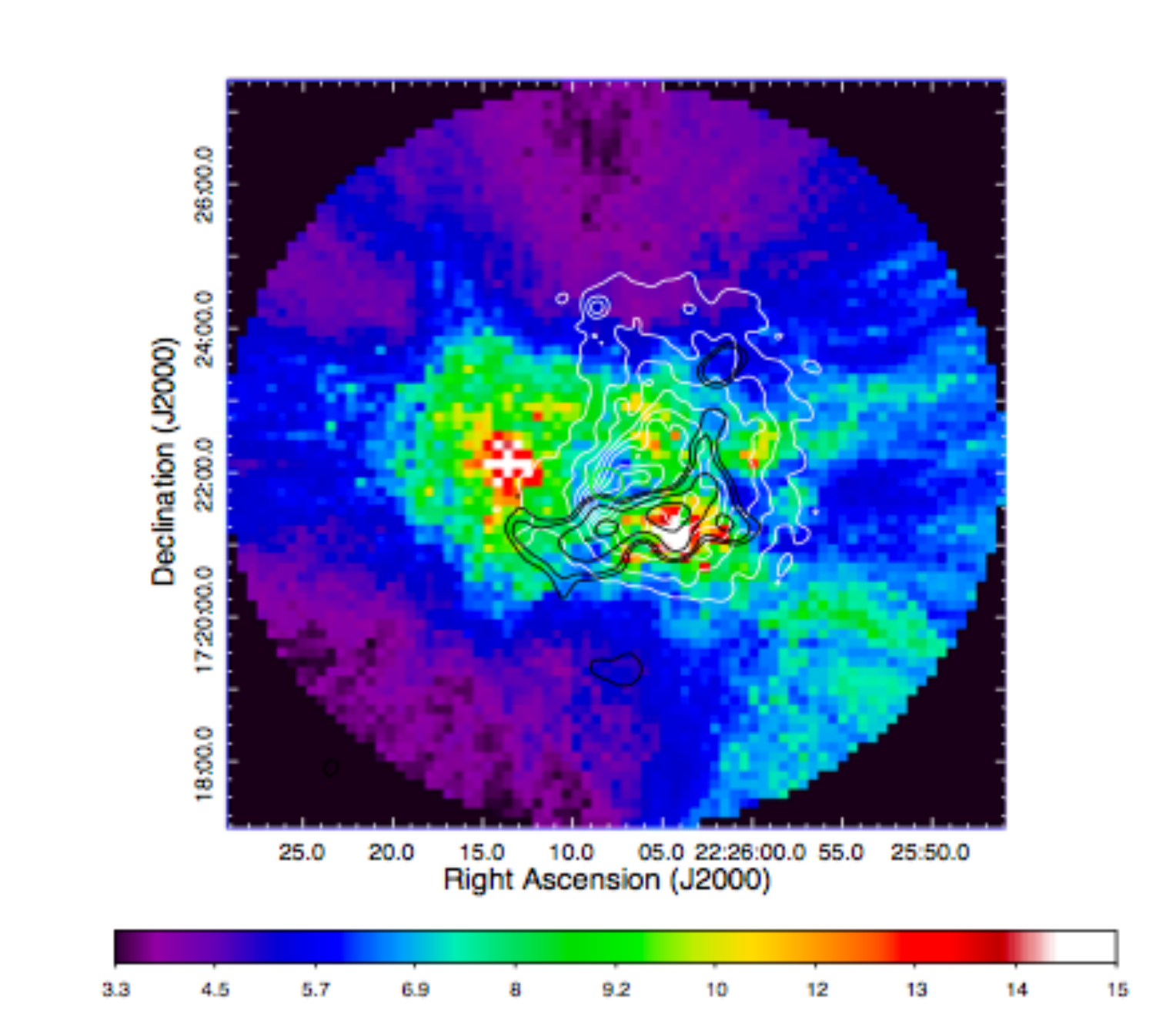} 
\includegraphics[width=0.3\textwidth]{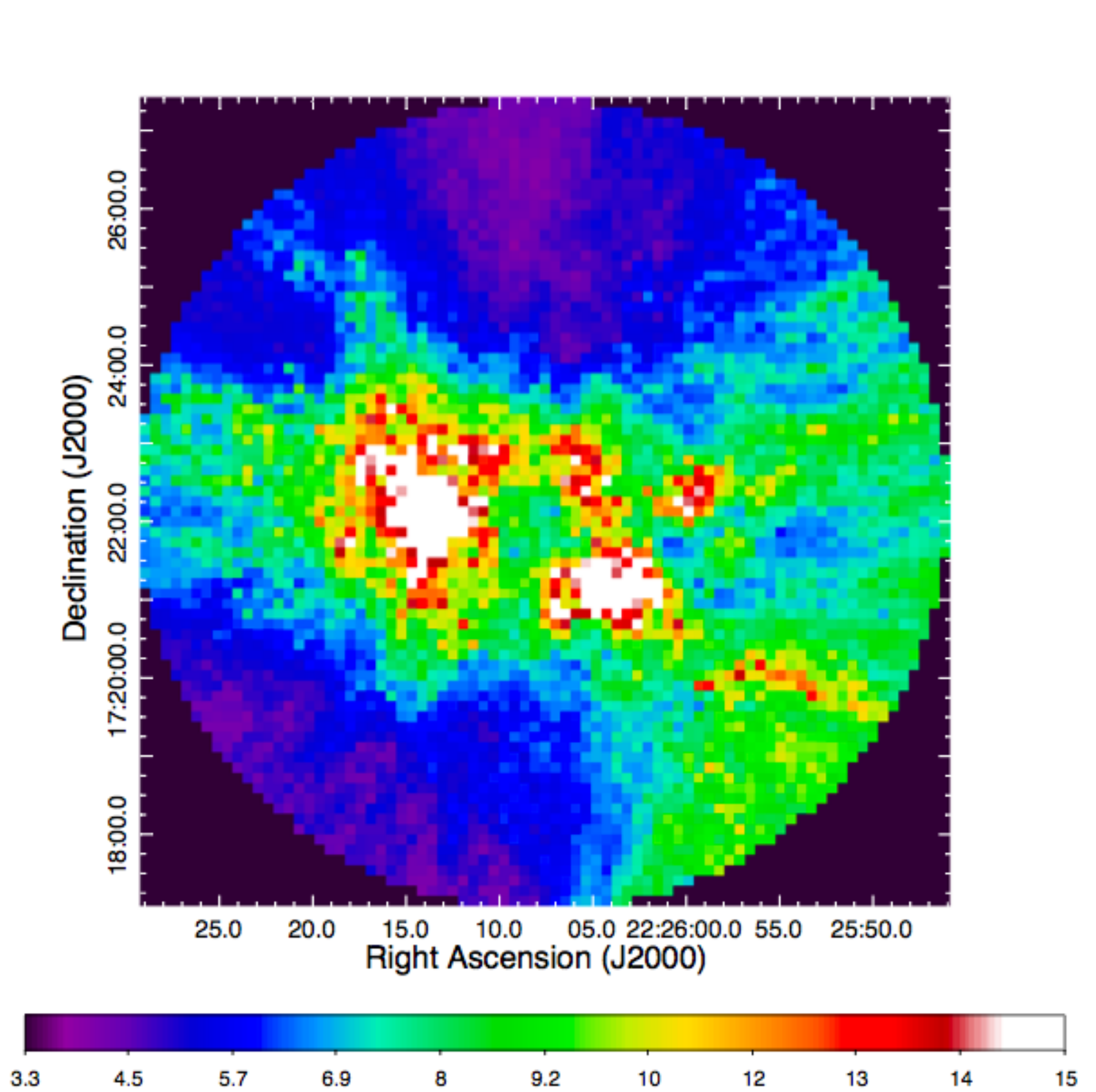}
\includegraphics[width=0.3\textwidth]{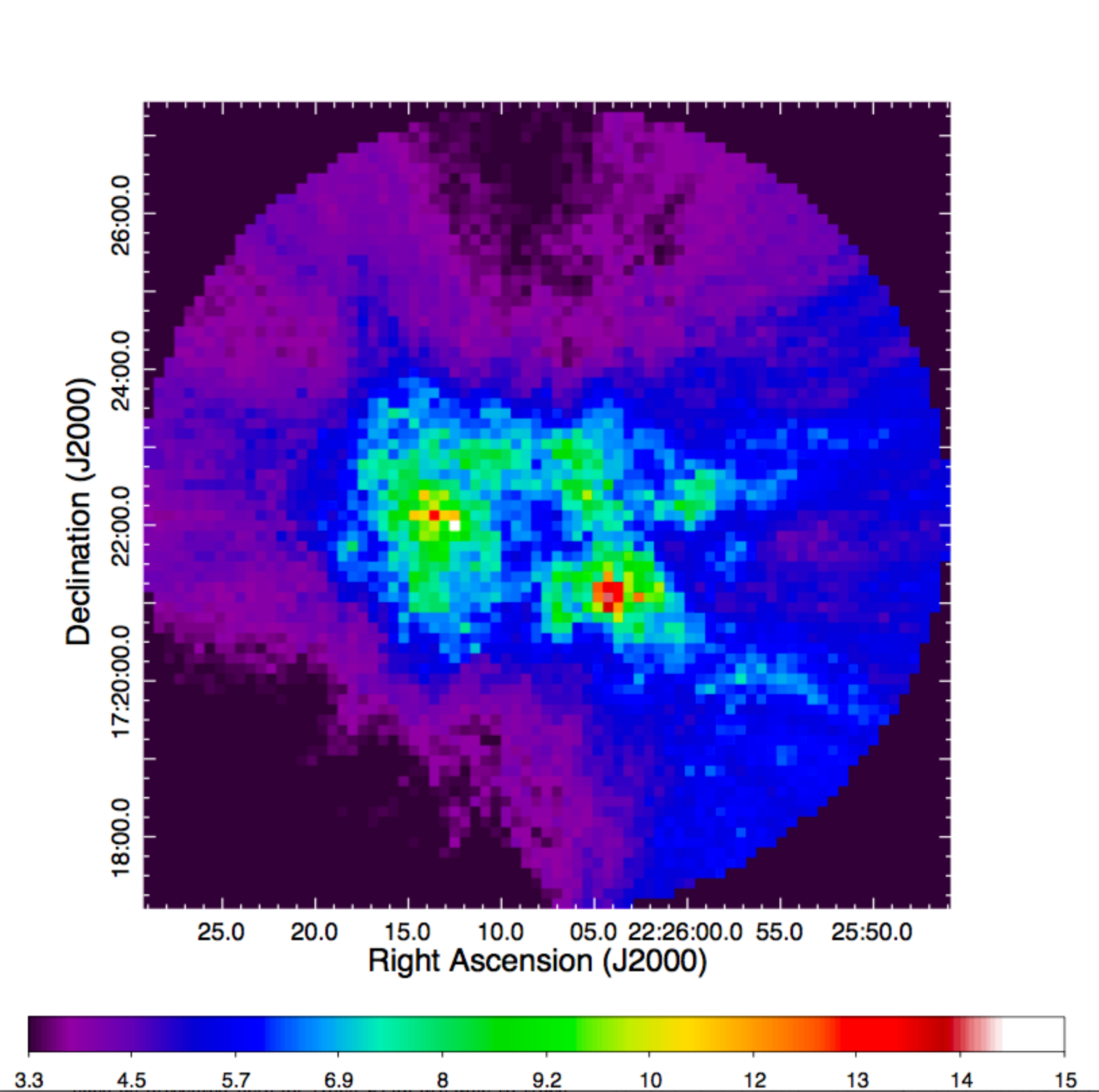}
\caption{\small {{\it Top Left:} $Chandra$ temperature map of Abell 2443
  showing two hot regions straddling a cooler region near the cluster
  center. The temperature at each pixel is determined from a spectral
  extraction region containing at least 1000 background subtracted
  counts. Colorbars for all plots show the temperature range in
  keV. VLA 74 MHz contours from \citet{cohen11} are shown in black and
  the smoothed $Chandra$ $0.3-7.0$ keV contours are shown in
  white. {\it Top Right:} Temperature error map (1$\sigma$) for the
  upper limits to the temperature fits. {\it Bottom Left:} Temperature
  error map (1$\sigma$) for the lower limits to the temperature fits.
\label{fig:tmap}}}
\end{figure*}

The intracluster medium in Abell 2443 shows evidence of an extended,
disturbed morphology, consistent with a dynamically complex merging
system. In order to look at the global cluster properties we have fit
the total intracluster emission (which covers the majority of the
ACIS-S3 CCD) using a circular region of radius 330 kpc centered on the
bright central emission. For all model fitting we have fixed the
absorption at the Galactic value. Using a single temperature absorbed
{\sc apec} model, we find a best fit temperature of
$kT=7.12^{+0.38}_{-0.33}$ keV and abundance\footnote{We report abundances
  relative to Solar abundances assuming the \citet{ag} abundance
  measurements.} of
$Z=0.37^{+0.09}_{-0.09}$ $Z_\odot$, with a reduced $\chi^2$ of $\chi^2_\nu = 227.9/201$.
Allowing for non-Solar abundances in the {\sc vapec} model results in
a reduced $\chi^2$ of $\chi^2_\nu = 223.8/200$ which is a marginally
significant improvement according to the $F$-test. The best-fit {\sc
  vapec} model gives $kT=7.00^{+0.38}_{-0.29}$ keV, and abundances of
Si=$1.35^{+0.50}_{-0.49}$ and Fe=$0.36^{+0.09}_{-0.09}$ with all other
elements constrained and fixed at 0.4. Deeper X-ray observations are
required to confirm and/or improve fits to non-Solar abundances.

\subsection{Temperature Map}
\label{sect:temp_map}

We have constructed a temperature map of Abell 2443 using the method
employed by \citet{randall08}. For each temperature map pixel, we
extracted a spectrum from a circular region containing roughly 1000
net counts after background subtraction.  The extraction regions
ranged in radius from about 55\arcsec\ ($\sim$ 107 kpc) in the bright
central regions to 238\arcsec\ ($\sim$ 464 kpc) in the faint outer
regions to the east. This mapping process results in extraction
regions that are significantly larger than the pixels shown in the
resulting temperature map, therefore the map pixels are not
independent, and the maps are effectively smoothed. For each
temperature map pixel, the resulting spectrum was fitted in the
$0.6-8.0$~keV range with an absorbed {\sc apec} model using {\sc xspec},
with the abundance allowed to vary.

The temperature map and associated 1$\sigma$ upper
and lower limits are shown in Figure~\ref{fig:tmap}. The main purpose
of the temperature map is to identify interesting features for further
study. Once features are identified, we fit extracted spectra of the
regions to determine the temperatures and assess the statistical
significance of the features.

The temperature map shows two distinct temperature peaks, one
67\arcsec\ ($\sim$ 130 kpc) southwest of the core, the other
82\arcsec\ ($\sim$ 160 kpc) east of the core.  We extracted spectra
from a 30~arcsec (58.5 kpc) radius region centered on the southwestern
temperature peak and from an identical, non-overlapping region just
northwest of the peak in a region of comparable surface brightness.
We fitted each spectrum with an absorbed {\sc apec} model, with the
abundance fixed at 30\% Solar.  We find temperatures of $kT_{\rm
  sw-peak} = 18.1^{+8.4}_{-5.1}$~keV for the temperature peak and
$kT_{\rm sw-offset} = 6.8^{+1.2}_{-0.9}$ keV for the adjacent region.  The
background makes up $< 6$\% of the total emission in this region, so
that the systematic uncertainties associated with the background do
not significantly affect the error budget.  Thus, the temperature
enhancement at the southwestern peak is significant at $\sim
2.2\sigma$.  The temperature peak is equally well-described by an
absorbed power-law model, with a photon index of $\Gamma =
1.40^{+0.06}_{-0.06}$.  Due to the low number of counts and the
implied high temperature of the thermal plasma, we were unable to
distinguish between these two models.

Performing a similar test for the eastern temperature peak by
extracting spectra from a 54~arcsec ($\sim$ 105 kpc) radius region
centered on the peak and from an identical, non-overlapping region
just south-southwest of the peak, we find temperatures of $kT_{\rm
  e-peak} = 20.9^{+29.4}_{-9.0}$~keV and $kT_{\rm e-offset} =
5.5^{+2.3}_{-1.4}$~keV.  This implies that the temperature enhancement
at the eastern peak is significant at $\sim 1.7\sigma$.  However, the
background makes up $\sim 30$\% of the total emission in these lower
surface brightness regions, so that systematic uncertainties in the
background contribute significantly to the total error budget.
Including a 5\% systematic uncertainty in the background
normalization, we find that $kT_{\rm e-peak} =
20.9^{+36.0}_{-10.0}$~keV and $kT_{\rm e-offset} =
5.5^{+3.2}_{-1.6}$~keV.  Thus, including the systematic uncertainty in
the background, the eastern temperature peak is significant at $\sim
1.5\sigma$.  As with the southwestern temperature peak, the spectrum is
equally well-modeled with an absorbed power-law model, with $\Gamma =
1.34^{+0.08}_{-0.08}$ (including the systematic background
uncertainty).
\begin{figure*}[t]
\includegraphics[width=0.475\textwidth]{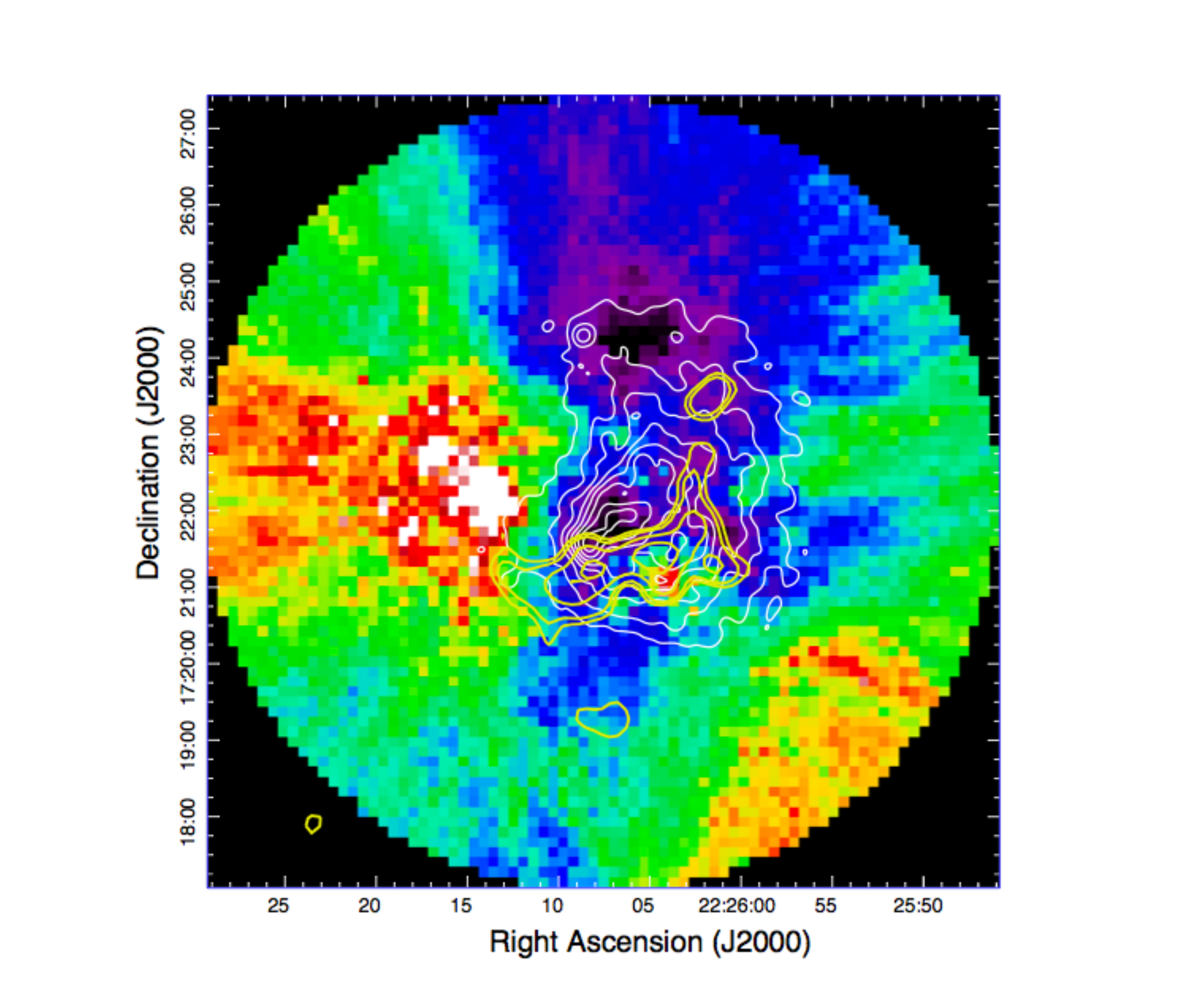} 
\hfill
\includegraphics[width=0.475\textwidth]{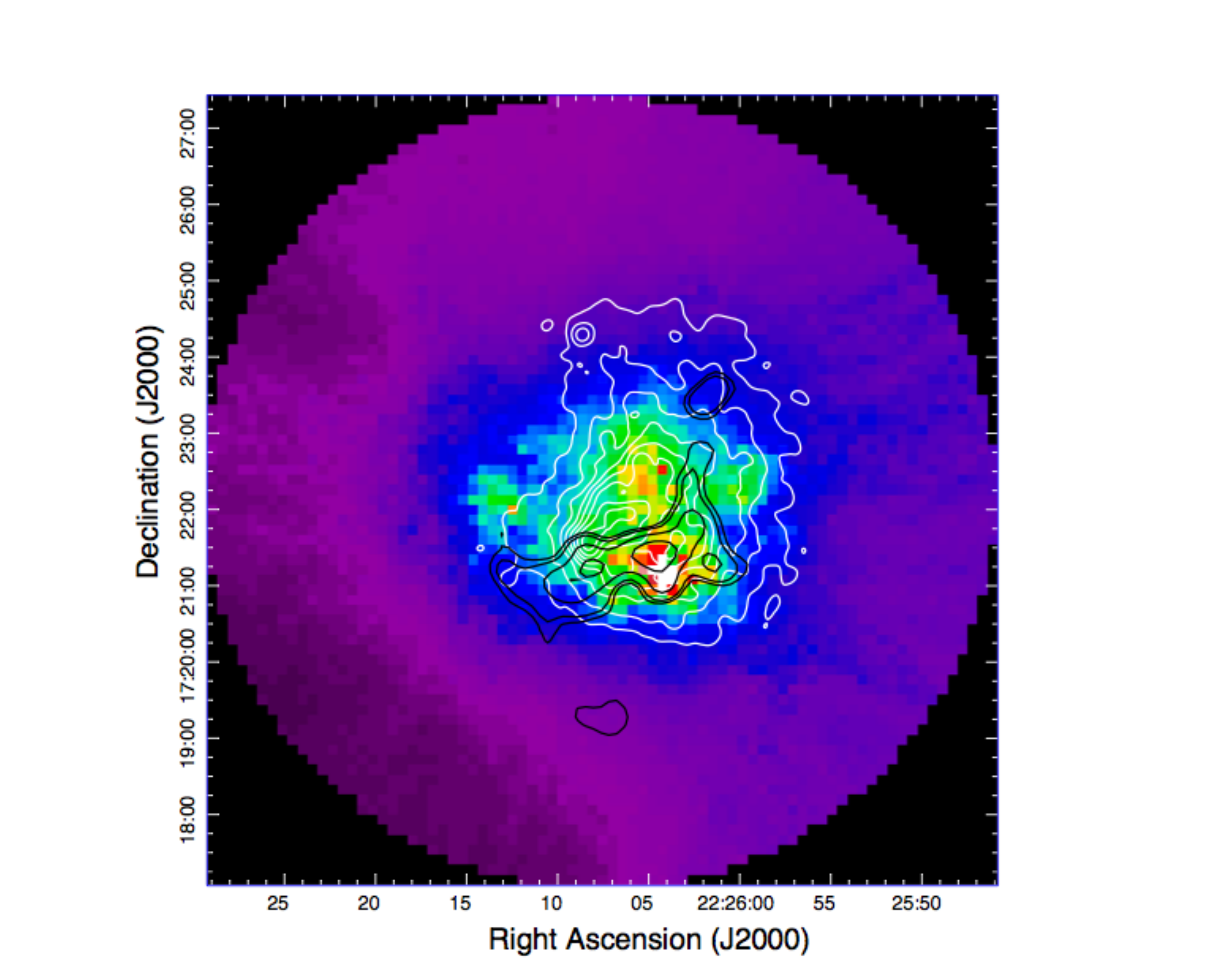} 
\caption{\small {The pseudo entropy (left) and pressure (right) maps, in
  arbitrary units (white/red regions are high and purple/black regions
  are low).  The entropy map was calculated as $kT A^{-1/3}$ and the
  pseudo pressure map as $kT A^{1/2}$, where $A$ is the {\sc apec}
  normalization scaled by the area of the extraction region and $kT$
  is the best fit temperature. Note that the radial extensions
    are due to smoothing artifacts in the maps that result from the
    fact that the extraction radii generally grow with distance from
    the bright central region. The entropy map shows low entropy gas
  associated with the cluster core and northern tail. The
  pseudo-pressure map shows a high pressure region near the relic
  which may indicate the presence of a shock. White contours show the
  smoothed $Chandra$ $0.3-7.0$ keV emission and the VLA 74 MHz
  contours from \citet{cohen11} are shown in yellow on the left panel
  and black on the right panel.
\label{fig:pentropy}}}
\end{figure*}

Since we cannot distinguish between thermal emission and non-thermal
power law emission in the regions of the high temperature peaks, a
concern is that unresolved sources are driving the hard spectra (and
thus, high temperatures) in these regions.  This is especially a
concern for the eastern temperature peak, where the X-ray surface
brightness is relatively low.  We therefore considered how many
unresolved sources would be required to drive the cooler temperature
measured adjacent to the eastern temperature peak ($kT_{\rm e-offset}
\approx 5.5$~keV) up to be consistent with the high peak temperature
($kT_{\rm e-peak} \approx 20.9$~keV) at 1$\sigma$ when fitting a
single temperature model by performing {\sc xspec} simulations.  We
assume background sources that are just below the 5$\sigma$ detection
limit, with a flat spectral index of $\Gamma = 0.3$, roughly the lower
limit of spectral indices observed in individual sources in deep blank
sky observations with {\it Chandra} \citep[e.g.][]{brandt01}.  These
are conservative assumptions, since the higher detection limit allows
for brighter unresolved sources, which will affect the blended spectra
more significantly, and flatter spectral indices contribute relatively
more photons at higher energies, also more significantly affecting the
blended spectrum.  We find that three such sources are required within
the 54~arcsec radius extraction region for the eastern temperature
peak to bring the temperature fitted for the blended spectrum into
agreement with the observed value, whereas the relationship between
the number of background sources $N$ above a given flux $S$ measured
from {\it Chandra} Deep Field observations \citep{brandt01} predicts
$\sim 0.3$ sources in the region of the eastern temperature peak.
Thus, while we cannot rule out the possibility of contamination from
unresolved point sources giving rise to the observed temperature
peaks, we conclude that this is unlikely. We discuss a possible
scenario for the hotspots in Section~\ref{sect:discussion}.

The temperature map also shows a cool plume of gas extending to the
north (Figure~\ref{fig:tmap}), in the region of the bright tail seen in
the X-ray image (Figure~\ref{fig:smo3}).  This feature is heavily
smoothed in the temperature map, owing to the low number of total net
counts in this region.  To test the significant of this feature, we
extracted spectra from a 75~arcsec by 47~arcsec ($\sim$ 146 $\times$
92 kpc) elliptical region covering the bright tail and from an
identical region in the fainter area northeast of the bright cluster
core (just north of the eastern temperature peak).  For the tail, we
find $kT_{\rm tail} = 4.8^{+0.5}_{-0.5}$~keV and an abundance of $Z_{\rm tail} =
0.3^{+0.2}_{-0.2}$ $Z_\odot$.  For the region east of the tail, we find $kT_{\rm
  tail-offset} = 8.4^{+4.8}_{-2.2}$~keV ($kT_{\rm tail-offset} =
8.4^{+6.9}_{-2.7}$~keV including a 5\% systematic uncertainty in the
background, as above).  The abundance could not be constrained in this
off-tail region, and was fixed at 30\% solar.  Thus, the cooler temperature in
the region of the tail is marginally significant, at 1.3$\sigma$.

We show in Figure~\ref{fig:pentropy} the pseudo-entropy and
pseudo-pressure maps for Abell 2443. The pseudo-pressure map shows
evidence of a jump near the location of the radio relic which would be
expected if this emission is associated with an X-ray shock. The main
features of the pseudo-entropy map are the low entropy gas associated
with the cooler cluster-center gas as well as a tail of low entropy
gas to the north that traces the northern X-ray tail seen in
Figure~\ref{fig:smo3}. The lack of a clear jump at the relic location
in the pseudo-entropy map is not surprising since the entropy jump for
a low to moderate Mach number shock is much less than the associated
pressure jump. Given the large extraction regions in the temperature
map, the entropy jump may be further masked by the blending of the
central cool gas into the shock region.

\subsection{Edge Analysis}
\label{sect:edge}

\begin{figure}
\includegraphics[width=0.55\textwidth]{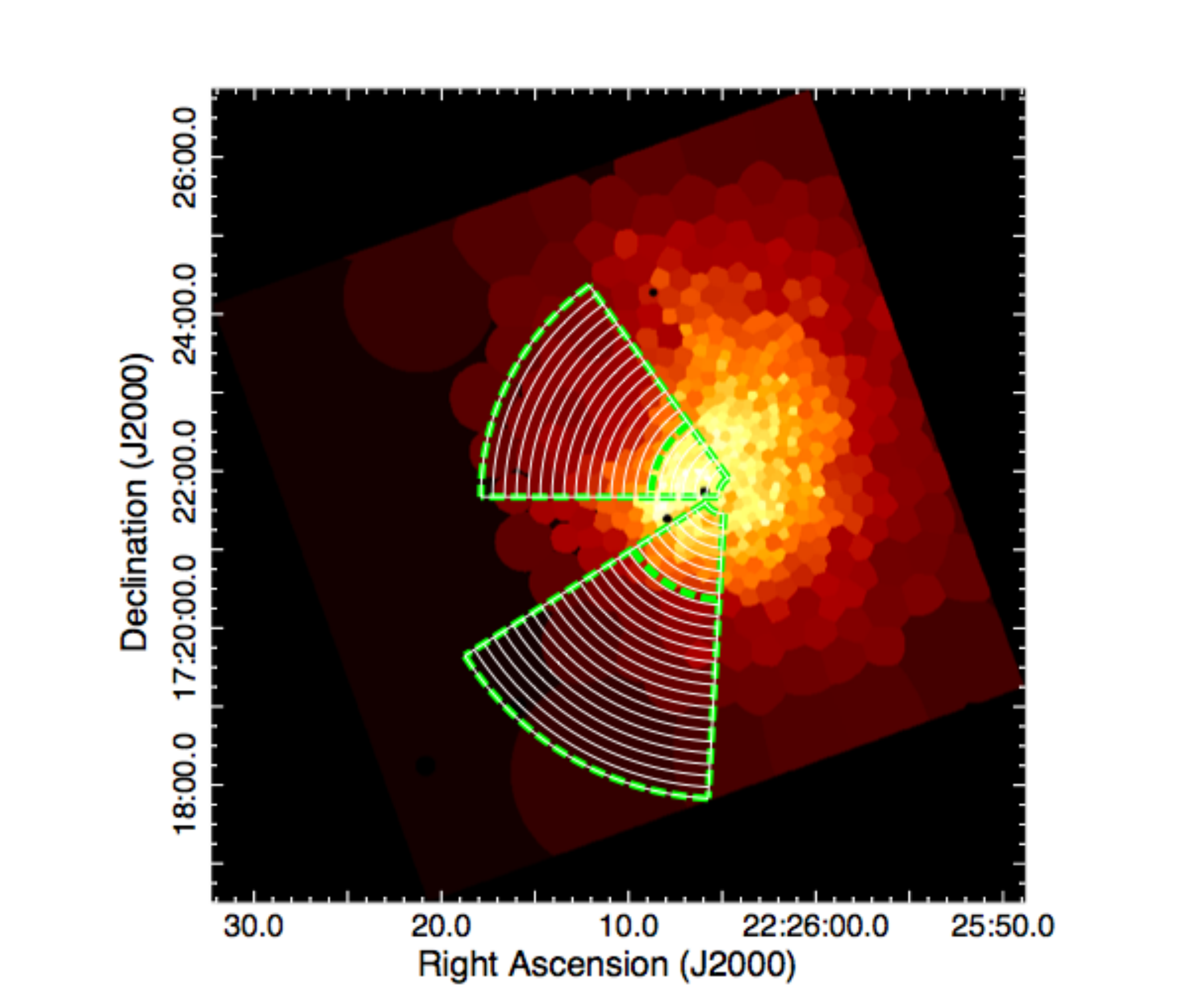}
\caption{\small {Voronoi tessellation image of the full ACIS-S3 CCD with the
  two surface brightness sectors overlaid in white. The inner and
  outer temperature wedges on either side of the two edges (discussed
  in \S~\ref{sect:edge}) are overlaid as dashed green regions.
\label{fig:vbin5_sect}}}
\end{figure}

Analysis of Figure~\ref{fig:smo3} reveals two possible X-ray edges,
one to the northeast and one to the southeast of the bright central
region. To characterize the apparent X-ray edges we have set two large
partial annuli (shown as dashed green regions in
Figure~\ref{fig:vbin5_sect}) on either side of the edges and fit the
temperature using an {\sc apec} model.  Results from this
  analysis suggest that the southeast edge may be a shock with inner
  hot shock-heated gas temperature of $kT_{\rm in}=10.9^{+4.2}_{-2.3}$
  and outer undisturbed gas temperature of $kT_{\rm
    out}=3.3^{+1.0}_{-0.6}$ keV. The northeast edge on the other hand
  has a cool inner temperatures of $kT_{\rm in}=6.5^{+0.9}_{-0.7}$ and
  hotter gas outside with $kT_{\rm out}=9.3^{+2.7}_{-1.6}$ keV. The
  presence of cool gas inside and hotter gas outside suggests that
  this may be a `cold front' edge rather than a shock. Cold front
  edges are contact discontinuities which are observed in both relaxed
  and merging clusters \citep[see review by][]{MV07}. We discuss the
  potential cold front further in Section~\ref{sect:tail} in
  connection with the northern X-ray tail.

\begin{figure}
\includegraphics[width=0.475\textwidth]{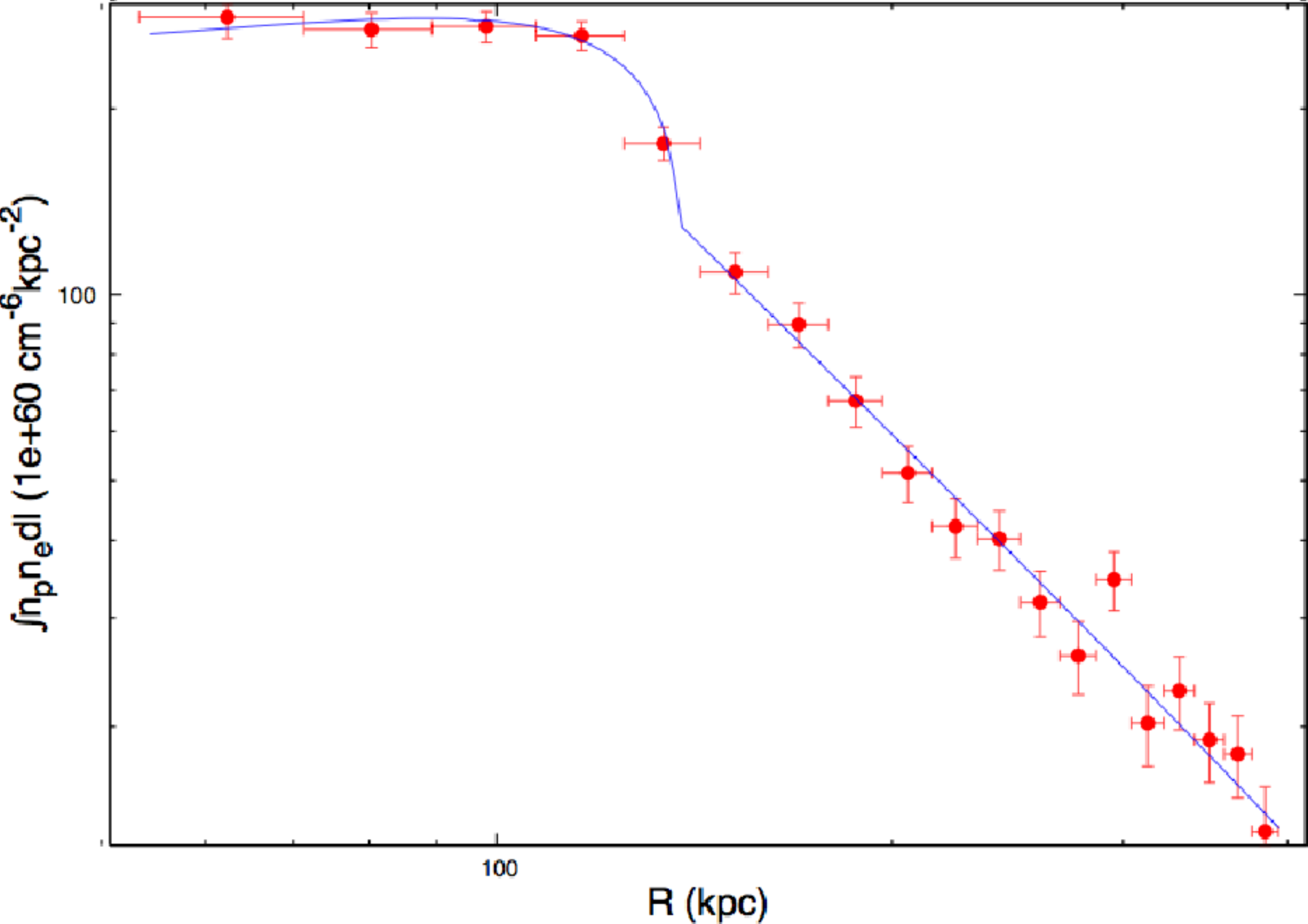}
\caption{\small {Edge fit to the X-ray surface brightness (data points) across
  the northeast `cold front' edge. The best-fit discontinuous
  power-law density model is shown as the curve.
\label{fig:nesect}}}
\end{figure}

\begin{figure}
\includegraphics[width=0.475\textwidth]{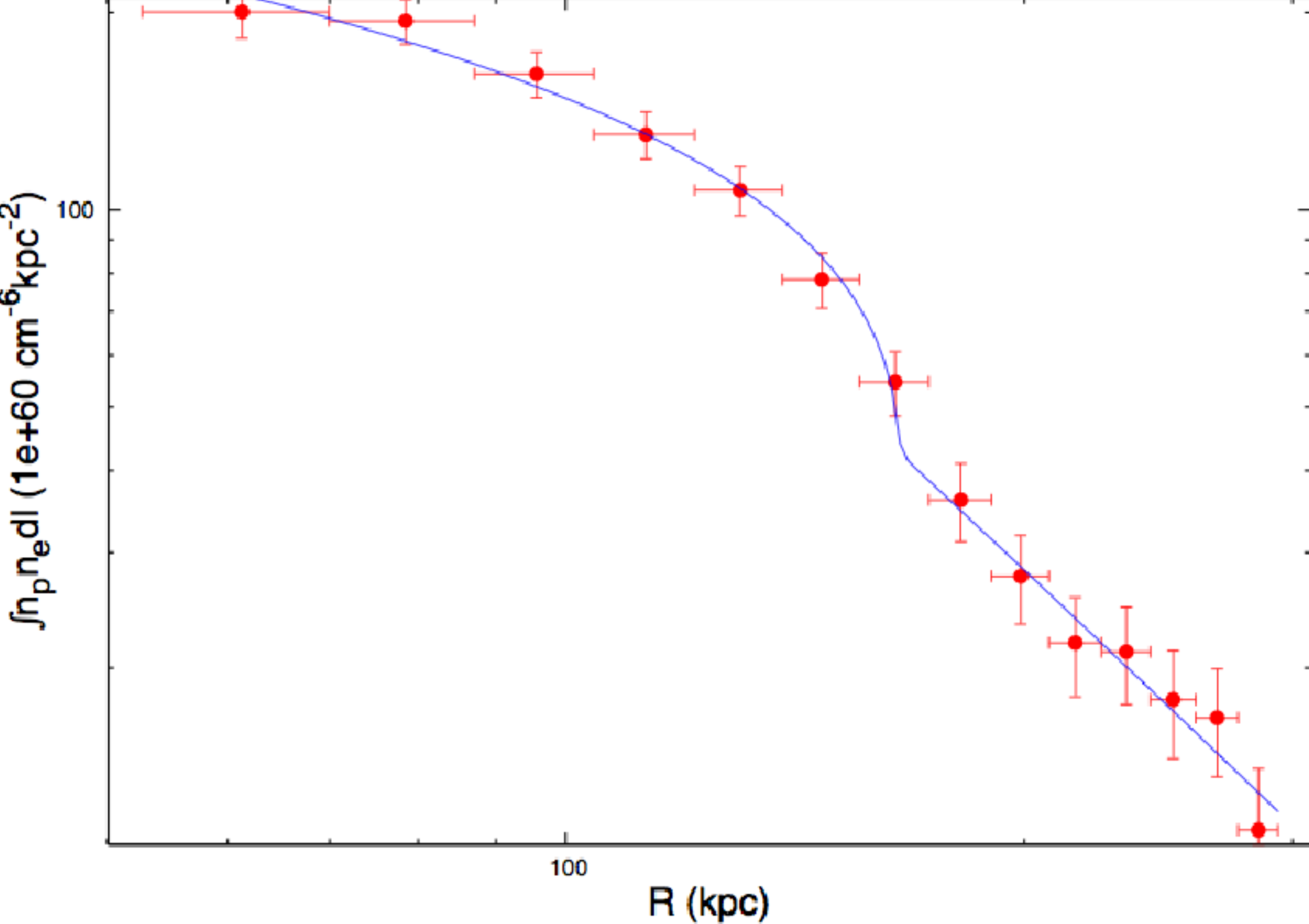}
\caption{\small {Edge fit to the X-ray surface brightness (data points) across
  the southeast `shock' edge. The best-fit
  discontinuous power-law density model is shown.
\label{fig:sesect}}}
\end{figure}

We have extracted surface brightness profiles across both edges using
the two pie-shaped sectors shown in Figure~\ref{fig:vbin5_sect}. The
sectors are defined to be centered on the center of curvature of the
X-ray edges rather than the ICM center or the BCG. The resulting
profiles are shown as the data points in Figures~\ref{fig:nesect}
\&~\ref{fig:sesect} for the northeast and southeast edges
respectively. The surface brightness profiles toward each edge were
fitted by projecting a 3D density model consisting of two power-laws
separated by a discontinuous jump as described in \citet{randall09}.
Due to the short exposure, there were insufficient counts to fit the
temperature or abundance profiles across the edges. Each surface
brightness profile was fit assuming $kT=7$ keV and an abundance of
$Z=0.5 Z_\odot$ that reflects an intermediate value between the core
and total abundance values. We note that using the pre- and post-edge
temperatures do not significantly change the calculated density jumps. 

The northeast `cold front' edge is consistent with a density jump of
$1.5-1.8$ and the best-fit model is shown as the
solid line in Figure~\ref{fig:nesect}. We note however that this model
fit has a non-physical positive density slope inside the edge which is
likely the result of applying a spherically symmetric model to this
complex system. The breakdown of the spherical symmetry assumption may
impact our fitting results but we note that the detection of the edge
is not impacted by this assumption. Deeper X-ray data are needed to
extract the profile in smaller bins near the edge to minimize the
effect of the spherical symmetry assumption. 

The southeast `shock' edge model fit is shown in
Figure~\ref{fig:sesect} and the associated best-fit density jump is in
the range of $1.2-1.8$. The best-fit density discontinuity of 1.6
would be consistent with a Mach number of ${\cal M} \approx1.4$
shock. The apparent temperature jump we measure above is a factor of
3.3 which would imply a stronger shock (${\cal M} \approx2.8$) than
measured by the density jump. Ideally we would like to measure the
temperature jump using small bins on either side of the edges in order
to clearly distinguish the jumps due to the edges from the underlying
large-scale trend with radius. Unfortunately we are only able to
measure the temperature jump in very large pie bins and thus there may
be large systematic errors in the measurements. Other effects such as
projection and non-equipartition could also reduce the observed
temperature jump. Given the large uncertainties in these measurements
due to the short exposure and low counts outside the bright core, the
Mach number is still very uncertain. Despite the uncertainty, both the
density and temperature measurements suggest the presence of a shock
near the outer edge of the USS radio source. A detailed analysis of
the edges to better constrain the parameters will require much deeper
$Chandra$ observations.

\section{Radio Spectral Analysis}
\label{sect:radio}

The Abell 2443 radio relic emission is very steep spectrum and shows
spectral curvature in the integrated spectrum
\citepalias{cohen11}. These spectral characteristics suggest that this
source is most likely the result of compression of fossil radio plasma
by a merger shock. The steep, curved spectral index is characteristic
of a plasma which is heavily affected by synchrotron and inverse
Compton losses \citep{ensslin01}.  Shock reacceleration can produce
steep spectrum emission but the spectrum is well-fit by a power-law
that connects the observed injection spectral index to the shock Mach
number \citep{pinzke}. We therefore suggest that the Abell 2443
  emission can be classified as an adiabatically compressed radio
  phoenix. Other examples of sources which fall within this class are
  the four USS sources studied by \citet{slee01}, as well as the more
  recent sources presented by \citet{vanw09a,vanw09b,vanw11},
  \citet{ogrean11}, and \citet{kale12}. The spatial co-incidence of
  the USS radio source with the possible X-ray shock in Abell 2443
  provides further support for the relic interpretation.

If the radio emission results from adiabatic compression of fossil
radio plasma, then the fossil plasma should not be older than 0.2 to
2.0 Gyr prior to the shock passage, depending on the external pressure
\citep{ensslin01}. Typical relics generated through adiabatic
compression are generally limited to scales of order 100 kpc or less
due to radiative energy losses sustained by the relic over the
timescale needed to compress a large system. Most relics are observed
at GHz frequencies where we would expect radiative losses to remove
the electrons responsible for the emission on time scales of a few
tens of Myr after shock passage for typical relic magnetic field
strengths. Observations at lower frequencies, such as presented here
for Abell 2443, are able to trace the relic emission from older
regions of compression. We might therefore expect to see both larger
adiabatically compressed relics as well as a spectral gradient across
the relic due to aging behind the region of current compression.

It is possible that the Abell 2443 relic is the result of a more
complex scenario such as a mix of compression and simultaneous
reacceleration of pre-existing relativistic particles. In order to
better characterize the relic we use the radio data of
\citetalias{cohen11} to investigate the spatial distribution of the
spectral index along the $\sim$ 350 kpc length the relic. We note that
we do not currently have sufficient spatial resolution across the
$\sim 100$ kpc width of the relic at multiple frequencies to look for
the spectral gradient along that direction.

We have divided the steep spectrum emission identified by
\citetalias{cohen11} as the relic region into three contiguous,
roughly equal area regions across the east-west axis of the
emission. We show these regions overlaid on the 74 MHz contours in
Figure~\ref{fig:vt_5_74}. We have clipped the 74 MHz A configuration
image and matched-resolution 330 MHz B configuration image of
\citetalias{cohen11} at the 5$\sigma$ level for the respective data
sets. Both data sets have been convolved to an identical circular beam
size of 23\farcs4. The middle of the three regions contains emission
from the compact source C while the other two regions (east and west)
are free of contamination from discrete sources. We have measured the
total flux within each of the three regions at 74 and 330 MHz and
corrected the flux measured in the middle region for the excess
emission from source C. Errors on the flux measurements for each
region include contributions from the rms noise as well as the
addition of a flux scale uncertainty of 3\% at 330 MHz and 6\% at 74
MHz. These flux measurements and uncertainties result in spectral
index measurements between 74 and 330 MHz of
$\alpha_{east}=-2.19\pm{0.07}$, $\alpha_{middle}=-1.72\pm{0.07}$, and
$\alpha_{west}=-1.59\pm{0.05}$. These values are consistent with a
spectral steepening from west to east that is significant at roughly
the 7$\sigma$ level. 

This spectral steepening eastward across the relic may be a reflection
of the original uncompressed spectrum that was more aged (steeper) on
the eastern edge compared to the western edge. Such a steepening could
be consistent, for example, with what would be expected for spectral
aging of plasma in the tail of source B where the plasma ages
eastward, away from the head. Additional spectral steepening would be
expected if the shock moved across the length of the relic from the
east toward the west. Alternatively, the spectral variation along the
length of the relic may be due to shock acceleration with a variation
of the Mach number along the relic, or due to a spectral gradient of a
pre-existing population of relativistic electrons that was
reaccelerated by the shock passage.

A detailed study of the spatial distribution of the spectrum and spectral
curvature will be presented in an upcoming paper that incorporates
additional new high-resolution, multi-frequency radio observations
from the Giant Metrewave Radio Telescope (GMRT) and the VLA.

\section{Discussion}
\label{sect:discussion}

We present an analysis of the first pointed X-ray observation of the
cluster Abell 2443 that is host to an USS radio source. Our new
$Chandra$ observation reveals that the ICM is elongated along a NW to
SE axis in the inner regions, and has an apparent X-ray tail to the
north in the outer regions. In addition, the X-ray surface brightness
shows two clear edges in the central region with an inner edge to the
NE and outer edge to the SE. In this section, we discuss the
implications of our analysis in more detail.

\subsection{Abell 2443 Environment}

The inner ICM elongation from NW to SE is similar to the optical
galaxy distribution of photometrically-determined cluster member
galaxies in a 2 Mpc region surrounding Abell 2443 \citep{wen07}. We
also note that there is a second galaxy cluster, ZwCl 2224.2+1651,
with associated X-ray emission from ROSAT, that is located roughly 2.7
Mpc SE of Abell 2443. Although this system has no spectroscopic
redshift measurements, there are numerous photometric redshifts of
galaxies within the region of this system that are consistent with the
redshift of Abell 2443 \citep{wen07}. In addition, there appears to be
a relatively higher fraction of blue galaxies within the SE cluster
which \citet{wen07} interpret as possibly resulting from recent star
formation triggered by the on-going interaction with Abell 2443. If
the systems are undergoing a merger it is not clear from the current
data what stage the merger is in. Based on the fact that there is a
concentration of galaxies associated with the X-ray emission for the
southern cluster, it may be in the early stages of
infall (possibly along a filament) since we would expect ram pressure
stripping of the X-ray emission from the southern system for any late
stage encounter with Abell 2443.

In order to further investigate the merger state of Abell 2443, we
have searched for redshift measurements of the galaxies in the
vicinity of the cluster. We find that there are four
spectroscopically-confirmed cluster-member ellipticals that are within
one magnitude in the $I-$band of the BCG identified by
\citet{crawford99}. These systems (identified in
Figure~\ref{fig:l_xray}) are located in a clump of three sources to
the NW of the BCG and a fourth elliptical to the NE of the BCG.

\subsection{Northern X-ray Tail}
\label{sect:tail}

The outer X-ray tail seen to the north in Abell 2443 arcs toward the
NE of the cluster and is reminiscent of the tails seen for example in
the infall of M86 into the Virgo Cluster \citep{randall08}, the early
stage cluster merger in Abell 115 \citep{A115} and the intermediate
stage cluster merger in Abell 1201 \citep{A1201}. These tails have
been interpreted as ram pressure stripped subhalos that consist of
material stripped from the host systems as it passes through a diffuse
external ICM.

We consider two possible ram pressure stripping scenarios for the
northern tail of Abell 2443. The first scenario we consider is the
case where the tail may be stripped material from the cluster as it
falls into the ICM of another system. We have searched the ROSAT
All-Sky images in the region near the cluster to try to identify a
potential stripping medium. The only nearby cluster is ZwCl
2224.2+1651, discussed above, which is likely associated with the
ROSAT Bright Source Catalog source 1RXSJ 222715.9+17035
\citep{RASSBSC}. At the projected distance of 2.7 Mpc, it seems unlikely
that the tail in Abell 2443 could be stripped material from the
interaction with the distant system.  If ZwCl 2224.2+1651 was
responsible for stripping Abell 2443, then the virial radius would
need to be nearly 3 Mpc, as would be expected only for very massive
clusters. Based on the X-ray emission, if the SE cluster is at the
same redshift as Abell 2443 then it must be significantly less massive
than Abell 2443 and thus could not be responsible for ram-pressure
stripping the Abell 2443 ICM. 

The second ram pressure stripping scenario we consider for the tail is
that Abell 2443 is a binary merger system that is currently in an
intermediate merger stage. If we consider two merging systems that
have roughly equal masses, then the northern tail may be at least
partially composed of stripped material from one or both of the
merging subhalos. An additional component of the tail could come from
merger-induced sloshing which is discussed below. Simulations of
equal-mass mergers shown in \citet{poole06} reveal evidence of
stripped X-ray tails with morphology very similar to Abell 2443. The
lack of two distinct X-ray cores in the cluster, together with the
X-ray core elongation suggest that the merger is most likely in an
intermediate stage, after both cores have disrupted, but before they
merge and relax. The presence of multiple bright elliptical galaxies
associated with Abell 2443 supports the intermediate stage binary
merger scenario.

An alternative scenario for the northern tail of Abell 2443 is that it
is an extended sloshing spiral resulting from the off-axis passage of
a perturbing subcluster. Hydrodynamic simulations show that
interactions with a passing subcluster can displace the cold central
gas from the minimum of the cluster potential. The cold gas will
overshoot the dark matter core as it falls back in, leading to the
sloshing effect \citep{slosh06}. Observationally, the X-ray signature
of a sloshing core is a large spiral-shaped plume of cool, high
metallicity gas running outward from the cluster core. Such sloshing
spirals are seen in numerous cool core systems (e.g.\ Abell 2029,
\citealt{clarke}; Abell 2052, \citealt{blanton11}; \citealt{lagana}). 

The majority of cool-core clusters with deep X-ray data show the
presence of `cold fronts', generated by sloshing, which trace the edges
of the sloshing spiral \citep{MV07}. We show that the northern tail of
Abell 2443 seen in Figure~\ref{fig:tmap} is associated with a cool
plume of gas although we caution that spectral fitting of the tail and
an off-tail region (discussed in \S~\ref{sect:temp_map}) indicate
that the cooler temperature in the tail region is only significant at
the 1.3$\sigma$ level. Nevertheless, if the outer edge of the tail in
Abell 2443 is the signature of a sloshing spiral, then the sloshing
spiral extent would be roughly 400 kpc. In this scenario, the inner
edge we detect to the NE of the core would simply be a cold front that
traces a portion of the inner region of the sloshing spiral near the
core. This interpretation of the edge is consistent with our
temperature estimates discussed in Section~\ref{sect:edge}. 

Numerous clusters show evidence of multiple cold fronts or sloshing
features that extend from a few 10's of kpc out to 200-300 kpc with
recent evidence in Perseus of an extent up to a Mpc
\citep{sim12}. Although our current X-ray data are not sensitive
enough to trace the full sloshing spiral (if it exists) we note that
the sloshing extent is similar to recent results from deep $Chandra$
observations of Abell 2029 which show a 400 kpc spiral, the largest
{\it continuous} sloshing spiral known to date \citep{A2029}. In the
case of Abell 2443, we may be viewing the spiral at some angle to the
line of sight (rather than face on) and thus would expect a more
complex X-ray signature than expected for face on, as shown in the
simulations of \citet{zuhone13}. We have tried standard
  techniques of unsharp-masking as well as fitting and
  subtracting/dividing an elliptical surface brightness model in order
  to enhance the residual surface brightness structures. Unfortunately
  these techniques do not produce additional insight into the
  structure, likely due to the short exposure. Deeper X-ray
  data on Abell 2443 are needed to search for additional cold fronts,
  better constrain the nature of the northern X-ray tail, and
  determine whether the tail has a clear signature of a sloshing
  spiral.

It is not clear what subhalo may be the perturber in the sloshing
spiral scenario although we note that it is not unusual that the
perturber is difficult to identify given the complexity of regions
around clusters and the expected disruption of the perturber by the
close encounter. Simulations of merger-induced sloshing by
\citet{zuhone} show that X-ray signatures persist for several Gyr
following closest core passage. Assuming relative motion in the plane
of the sky of 2000 km/s between Abell 2443 and ZwCl 2224.2+1651, the
two systems would have had closest approach roughly 1.3 Gyr ago, well
within the timescale over which sloshing cores persist. 

\subsection{Surface Brightness Edges}

In Section~\ref{sect:edge} we discuss the details of the two surface
brightness edges we detect within Abell 2443. As discussed above, the
inner edge to the NE may be a cold front edge. Analysis of the
temperature in large bins on either side of the edge shows cooler
temperatures inside the edge and hotter temperatures outside. We note
however that while the surface brightness profile shows a clear edge,
we do not have sufficient counts to trace the temperature or pressure
profiles in detail across the edge. Our model fits assume spherical
symmetry, and we find a non-physical positive density slope inside the
edge, therefore, future work on deeper X-ray data may require a more complex
geometry to model the edge.

The outer edge to the SE is consistent with a shock, with inner
temperatures higher than outer temperatures. The X-ray surface
brightness has a best-fit density jump in the range of $1.2-1.8$
(90\% confidence). Although the density jump is consistent with a Mach
number of ${\cal M} \approx1.4$ shock, the measured temperature jump
would imply a much stronger shock.  Unfortunately, the temperature
measurements had to be performed in very wide sectors on either side
of the edge and are therefore not necessarily representative of the
jump across the edge. In particular, with the current data we cannot
distinguish a coarsely sampled temperature profile that smoothly drops
with radius from a sharp jump across a shock edge. Deeper X-ray data
will be required to trace the temperature on finer spatial
scales. Confirmation of the merger shock would support the
intermediate stage binary merger as we would not expect to detect a
shock in a sloshing system that was driven only by a close encounter.

In principle the shock Mach number could be much higher than we
estimate. If the shock is really a Mach 3-4 shock that we are viewing
at an angle, then we would measure a much lower temperature
and pressure jump due to projection effects. In addition, the density
jump can also be underestimated for such a geometry if we are looking
along a tangent to the Mach cone (i.e.\ the merger axis is not close
to the plane of the sky). 

\subsection{Temperature Map}

The X-ray temperature map in Figure~\ref{fig:tmap} shows two hot
regions located to the NE and SW of the cooler cluster core. We find
that it is unlikely that these apparent hot regions could be driven by
faint unresolved sources just below our $Chandra$ detection limit.

One plausible interpretation for these two hot regions is that they
are tracing the shock Mach cone which is being seen at some angle to
the plane of the sky such that we are looking through the Mach cone.
In this case, we would be seeing the cooler cluster core dominating
the emission in the central region where the Mach cone is
superimposed, and the hot regions of the Mach cone dominating in the
outer, fainter regions on either side of the core. Unfortunately it
would be very difficult to spectrally separate a cool cluster core
component and hot shocked gas component even with very deep X-ray
data. 

\subsection{Radio/X-ray Connection}

The spatial coincidence of the USS radio emission with the SE shock
edge supports our interpretation of the radio emission as a radio
relic. The elongated nature of the emission and non-central location
are also typical of radio relics \citep{ferrari08}. The indication of
a curved radio spectrum presented in \citet{cohen11} would suggest
that the source may be an adiabatically compressed fossil radio
lobe. Diffusive shock acceleration is able to produce USS relics but
their spectrum would be expected to be power-law. Currently we only
have flux measurement at three radio frequencies, thus we need to
expand the radio frequency measurements to more clearly differentiate
adiabatic compression from shock reacceleration of a pre-existing
relativistic particle population.  We note that in the shock
reacceleration scenario, the low frequency integrated spectral index
measured by \citetalias{cohen11} of $\alpha_{74}^{325} = -1.7$ would
be consistent with a ${\cal M} \approx2.0$ shock which is plausible
given our large uncertainties. We will explore the detailed spectral
properties further in a future paper.

In the adiabatic compression model, the most likely candidate for the
fossil plasma is the head-tail galaxy B. We have extracted the flux at
74 MHz and 330 MHz in three large, independent regions running from
west to east along the length of the relic in order to estimate the
variation of the spectral index along the source. We find that there
is evidence of spectral steepening from west to east that is
significant at the 7$\sigma$ level. This steepening could be due to
spectral aging of the relativistic plasma away from the head of source
B. If the entire tail is active and/or revived plasma from source B,
then we would be tracing the emission over a total arc length of 425
kpc from the head. We note, however, that the spectral variation we
measure could also be produced by a variation in the Mach number along
the shock edge, or by a spectral gradient in a pre-existing population
of relativistic electrons that was reaccelerated by the shock passage.

Finally, we consider the possibility that the USS source may instead
be a dying/dead radio galaxy. Integrated spectral measurements of
dying radio galaxies show that they can be characterized by very
steep, curved spectra \citep{murgia11}, similar to that of the Abell
2443 USS source. We note, however, that the dying radio galaxies
studied by \citeauthor{murgia11} retain knowledge of their original
spectrum such that spectral mapping shows the cores are flatter and
the source steepens toward the outer edges. In the case of the USS
source in Abell 2443, we see a spectral gradient with steepening from
west to east rather than evidence of symmetric steepening from the
center. This observed spectral signature does not appear to be
consistent with the dying radio galaxy scenario. Deeper
multi-frequency radio data will allow us to map the detailed spatial
distribution of the spectral index and curvature and will help better
distinguish these models.

\section{Summary}
\label{sect:summary} 

Our new $Chandra$ observation shows the presence of a disturbed ICM in
Abell 2443 which is suggestive of a merging system. The ICM is
elongated along a northwest to southeast direction, similar to the
photometric cluster member distribution discussed by
\citet{wen07}. The central region of the cluster shows two X-ray edges
which we tentatively identify as an inner cold front edge to the
northeast of the core and an outer shock edge to the southeast of the
core. There is also a cool X-ray tail seen to the north which may be
ram pressure stripped material and/or a sloshing spiral. Based on the
current data, we favor the model that the X-ray features are driven by
an intermediate stage binary merger. The presence of multiple bright
elliptical galaxies near the core supports the merger scenario.

The diffuse USS radio emission detected by \citetalias{cohen11} is
co-incident with the southeast shock edge; thus the non-central
location and connection to a potential shock strengthens the
identification of the emission as a radio relic.  Based on the density
jump at the southeast edge, we find that it is consistent with a Mach
number $M\simeq$ 1.4 shock, although the Mach number is still very
uncertain due to the large uncertainties in the measurements. We
interpret the two hot regions detected on either side of the central
core as areas where we are viewing the Mach cone emission dominating
the spectral fits while the cool core dominates in the center.

The steep spectral index and spectral curvature of the relic are
suggestive of adiabatic compression of fossil radio plasma (possibly
from the tail of radio galaxy B) as the origin of the relic. The size
of the relic is slightly larger than expected for this model. With
just a single high frequency flux measurement indicating spectral
curvature, we do not rule out the possibility that the USS relic is
due to shock reacceleration of fossil relativistic electrons. We have
investigated the spatial distribution of the radio spectral index in
three separate regions across the relic and find evidence of
significant spectral steepening from the western edge to the eastern
edge of the system. This spectral steepening would be consistent with
what would be expected for spectral aging along the tail of source B.

\section{Acknowledgments}

TEC was supported in part for this work by the National Aeronautics
and Space Administration, through Chandra Award Number
GO1-12007Z. Basic research in radio astronomy at the Naval Research
Laboratory is supported by 6.1 Base funding.  SR is supported by the
Chandra X-ray Center through NASA contract NAS8-03060 and the
Smithsonian Institution. CLS was supported in part by NASA Chandra
Grants GO9-0135X, GO9-0148X, and GO1-12169X, and NASA ADAP grant
NNX11AD15G. ELB was partially supported by NASA through the
Astrophysics Data Analysis Program, grant number NNX10AC98G, and
through NASA award RSA No. 1440385 issued by
JPL/Caltech. S.G.\ acknowledges the support of NASA through Einstein
Postdoctoral Fellowship PF0–110071 awarded by the Chandra X–ray
Center. The National Radio Astronomy Observatory is a facility of the
National Science Foundation operated under cooperative agreement by
Associated Universities, Inc. This research has made use of the
NASA/IPAC Extragalactic Database (NED) which is operated by the Jet
Propulsion Laboratory, California Institute of Technology, under
contract with the National Aeronautics and Space Administration.

Funding for SDSS-III has been provided by the Alfred P. Sloan
Foundation, the Participating Institutions, the National Science
Foundation, and the U.S. Department of Energy Office of Science. The
SDSS-III web site is http://www.sdss3.org/.

SDSS-III is managed by the Astrophysical Research Consortium for the
Participating Institutions of the SDSS-III Collaboration including the
University of Arizona, the Brazilian Participation Group, Brookhaven
National Laboratory, University of Cambridge, Carnegie Mellon
University, University of Florida, the French Participation Group, the
German Participation Group, Harvard University, the Instituto de
Astrofisica de Canarias, the Michigan State/Notre Dame/JINA
Participation Group, Johns Hopkins University, Lawrence Berkeley
National Laboratory, Max Planck Institute for Astrophysics, Max Planck
Institute for Extraterrestrial Physics, New Mexico State University,
New York University, Ohio State University, Pennsylvania State
University, University of Portsmouth, Princeton University, the
Spanish Participation Group, University of Tokyo, University of Utah,
Vanderbilt University, University of Virginia, University of
Washington, and Yale University.  

We thank the anonymous referee for helpful comments.

\begin{deluxetable*}{crrcccc}

\tablecaption{$Chandra$ Compact X-ray Sources
\label{tab:compact}}
\tablehead{
\colhead{Source} & 
\colhead{RA} & 
\colhead{Dec} & 
\colhead{redshift} & 
\colhead{Optical ID} &
\colhead{Radio ID} &
\colhead{XAssist Flux}\\
\colhead{} &
\colhead{(J2000)} &
\colhead{(J2000)} &
\colhead{} &
\colhead{} &
\colhead{} &
\colhead{(erg s$^{-1}$ cm$^{-2}$)}
}
\startdata
1 & 22 26 07.98 & +17 21 23.47 & 0.107 & BCG & C & 2.20$\times 10^{-14}$\\
2 & 22 26 05.96 & +17 21 44.67 & 0.227 & T96 & J & 1.75$\times 10^{-14}$\\
3\footnote{Source 3 is the lowest significance source (2.96) from {\sc wavdetect}.}  & 22 26 02.75 & +17 22 27.36 & 0.115 & T06 & B tail & --- \\
4 & 22 26 08.66 & +17 24 17.15 & 0.115 & --- & --- & 2.22$\times 10^{-14}$\\
5 & 22 25 58.48 & +17 21 25.00 & --- & star & --- & --- \\
6 & 22 25 54.43 & +17 23 10.72 & --- & A-star & --- & 7.62$\times 10^{-15}$
\enddata
\tablecomments{Column 1 lists the source ID number, columns 2 and 3
  are the right ascension and declination in J2000 measured from
  $Chandra$, column 4 is the source photometric redshift (if known), column 5 is the optical ID
  where BCG indicates the brightest cluster galaxy as indicated by \citet{crawford99} and 
  TNN refers to galaxy number NN in the catalog of
  \citet{trujillo01}, column 6 is the associated radio source ID from \citet{cohen11} or this paper, and 
  column 7 is the XAssist flux in the 0.3 to 8.0 keV band \citep{xassist}.}
\label{tbl:compact}
\end{deluxetable*}


\begin{thebibliography}{thisisasampleofwhat}

\bibitem[Anders \& Grevesse(1989)]{ag} Anders, E., \& Grevesse, N.\ 1989, \gca, 53, 197 

\bibitem[Ascasibar \& Markevitch(2006)]{slosh06} Ascasibar, Y., \& Markevitch, M.\ 2006, \apj, 650, 102 

\bibitem[Blanton et al.(2011)]{blanton11} Blanton, E.~L., Randall, S.~W., Clarke, T.~E., et al.\ 2011, \apj, 737, 99

\bibitem[Bliton et al.(1998)]{bliton98} Bliton, M., Rizza, E., Burns, J.~O., Owen, F.~N., \& Ledlow, M.~J.\ 1998, \mnras, 301, 609

\bibitem[Brandt et al.(2001)]{brandt01} Brandt, W.~N., Alexander, D.~M., Hornschemeier, A.~E., et al.\ 2001, \aj, 122, 2810

\bibitem[Bourdin et al.(2013)]{bourdin13} Bourdin, H., Mazzotta, P., Markevitch, M., Giacintucci, S., \& Brunetti, G.\ 2013, \apj, 764, 82

\bibitem[Cassano(2010)]{cassano2010} Cassano, R.\ 2010, \aap, 517, A10

\bibitem[Clarke et al.(2004)]{clarke} Clarke, T.~E., Blanton, E.~L., \& Sarazin, C.~L.\ 2004, \apj, 616, 178 

\bibitem[Crawford et al.(1999)]{crawford99} Crawford, C.~S., Allen, S.~W., Ebeling, H., Edge, A.~C., \& Fabian, A.~C.\ 1999, \mnras, 306, 857

\bibitem[Cohen \& Clarke(2011)]{cohen11} Cohen, A.~S., \& Clarke, T.~E.\ 2011, \aj, 141, 149

\bibitem[Cohen et al.(2007)]{cohen07} Cohen, A.~S., Lane, W.~M., Cotton, W.~D., Kassim, N.~E., Lazio, T.~J.~W., Perley, R.~A., 
Condon, J.~J., \& Erickson, W.~C.\ 2007, \aj, 134, 1245 

\bibitem[Dennison(1980)]{dennison80} Dennison, B.\ 1980, \apjl, 239, L93 

\bibitem[Dickey \& Lockman(1990)]{dl} Dickey, J.~M., \& Lockman, F.~J.\ 1990, \araa, 28, 215

\bibitem[Diehl \& Statler(2006)]{ds06} Diehl, S., \& Statler, T.~S.\ 2006, \mnras, 368, 497 

\bibitem[Ensslin et al.(1998)]{ensslin98} Ensslin, T.~A., Biermann, P.~L., Klein, U., \& Kohle, S.\ 1998, \aap, 332, 395 

\bibitem[En{\ss}lin \& Gopal-Krishna(2001)]{ensslin01} En{\ss}lin, T.~A., \& Gopal-Krishna 2001, \aap, 366, 26 

\bibitem[Feretti et al.(2012)]{feretti2012} Feretti, L., Giovannini, G., Govoni, F., \& Murgia, M.\ 2012, \aapr, 20, 54 

\bibitem[Ferrari et al.(2008)]{ferrari08} Ferrari, C., Govoni, F., Schindler, S., Bykov, A.~M., \& Rephaeli, Y.\ 2008, \ssr, 134, 93 

\bibitem[Finoguenov et al.(2010)]{fino10} Finoguenov, A., Sarazin, C.~L., Nakazawa, K., Wik, D.~R., \& Clarke, T.~E.\ 2010, \apj, 715, 1143 

\bibitem[Freeman et al.(2002)]{freeman02} Freeman, P.~E., Kashyap, V., Rosner, R., \& Lamb, D.~Q.\ 2002, \apjs, 138, 185

\bibitem[Gutierrez \& Krawczynski(2005)]{A115} Gutierrez, K., \& Krawczynski, H.\ 2005, \apj, 619, 161  

\bibitem[Kale \& Dwarakanath(2012)]{kale12} Kale, R., \& Dwarakanath, K.~S.\ 2012, \apj, 744, 46

\bibitem[Kang \& Ryu(2011)]{kang11} Kang, H., \& Ryu, D.\ 2011, \apj, 734, 18 

\bibitem[Kang et al.(2012)]{kang12} Kang, H., Ryu, D., \& Jones, T.~W.\ 2012, \apj, 756, 97

\bibitem[Keshet(2010)]{keshet10} Keshet, U.\ 2010, arXiv:1011.0729

\bibitem[Korngut et al.(2011)]{korgut11} Korngut, P.~M., Dicker, S.~R., Reese, E.~D., et al.\ 2011, \apj, 734, 10 

\bibitem[Lagan{\'a} et al.(2010)]{lagana} Lagan{\'a}, T.~F., Andrade-Santos, F., \& Lima Neto, G.~B.\ 2010, \aap, 511, A15 

\bibitem[Ma et al.(2012)]{A1201} Ma, C.-J., Owers, M., Nulsen, P.~E.~J., et al.\ 2012, \apj, 752, 139 

\bibitem[Macario et al.(2011)]{macario11} Macario, G., Markevitch, M., Giacintucci, S., et al.\ 2011, \apj, 728, 82

\bibitem[Markevitch et al.(2002)]{markevitch02} Markevitch, M., Gonzalez, A.~H., David, L., et al.\ 2002, \apjl, 567, L27 

\bibitem[Markevitch et al.(2005)]{markevitch05} Markevitch, M., Govoni, F., Brunetti, G., \& Jerius, D.\ 2005, \apj, 627, 733 

\bibitem[Markevitch \& Vikhlinin(2007)]{MV07} Markevitch, M., \& Vikhlinin, A.\ 2007, \physrep, 443, 1

\bibitem[Mazzotta et al.(2011)]{mazzotta11} Mazzotta, P., Bourdin, H., Giacintucci, S., Markevitch, M., \& Venturi, T.\ 2011, \memsai, 82, 495 

\bibitem[Miller et al.(2002)]{miller02} Miller, C.~J., Krughoff, K.~S., Batuski, D.~J., \& Hill, J.~M.\ 2002, \aj, 124, 1918

\bibitem[Murgia et al.(2011)]{murgia11} Murgia, M., Parma, P., Mack, K.-H., et al.\ 2011, \aap, 526, A148

\bibitem[Ogrean et al.(2011)]{ogrean11} Ogrean, G.~A., Br{\"u}ggen, M., van Weeren, R., et al.\ 2011, \mnras, 414, 1175 

\bibitem[Owers et al.(2011)]{owers11} Owers, M.~S., Randall, S.~W., Nulsen, P.~E.~J., et al.\ 2011, \apj, 728, 27 

\bibitem[Paterno-Mahler et al.(2013)]{A2029} Paterno-Mahler, R., Blanton, E.\ L., Randall, S.\ W., \& Clarke, T.\ E., 2013, ApJ submitted

\bibitem[Petrosian \& Bykov(2008)]{petrosian08} Petrosian, V., \& Bykov, A.~M.\ 2008, \ssr, 134, 207 

\bibitem[Pinzke et al.(2013)]{pinzke} Pinzke, A., Oh, S.~P., \& Pfrommer, C.\ 2013, arXiv:1301.5644

\bibitem[Poole et al.(2006)]{poole06} Poole, G.~B., Fardal, M.~A., Babul, A., et al.\ 2006, \mnras, 373, 881 

\bibitem[Ptak \& Griffiths(2003)]{xassist} Ptak, A., \& Griffiths, R.\ 2003, Astronomical Data Analysis Software and Systems XII, 295, 465 

\bibitem[Randall et al.(2009)]{randall09} Randall, S.~W., Jones, C., Markevitch, M., et al.\ 2009, \apj, 700, 1404 

\bibitem[Randall et al.(2008)]{randall08} Randall, S., Nulsen, P., Forman, W.~R., et al.\ 2008, \apj, 688, 208 

\bibitem[Russell et al.(2010)]{russell10} Russell, H.~R., Sanders, J.~S., Fabian, A.~C., et al.\ 2010, \mnras, 406, 1721

\bibitem[Sarazin(2000)]{sarazin00} Sarazin, C.~L.\ 2000, Constructing the Universe with Clusters of Galaxies,

\bibitem[Simionescu et al.(2012)]{sim12} Simionescu, A., Werner, N., Urban, O., et al.\ 2012, \apj, 757, 182

\bibitem[Slee et al.(2001)]{slee01} Slee, O.~B., Roy, A.~L., Murgia, M., Andernach, H., \& Ehle, M.\ 2001, \aj, 122, 1172 

\bibitem[Slee et al.(1994)]{slee94} Slee, O.~B., Roy, A.~L., \& Savage, A.\ 1994, Australian Journal of Physics, 47, 145 

\bibitem[Struble \& Rood(1999)]{sr99} Struble, M.~F., \& Rood, H.~J.\ 1999, \apjs, 125, 35 

\bibitem[Truemper(1993)]{RASS} Truemper, J.\ 1993, Science, 260, 1769 

\bibitem[Trujillo et al.(2001)]{trujillo01} Trujillo, I., Aguerri, J.~A.~L., Guti{\'e}rrez, C.~M., \& Cepa, J.\ 2001, \aj, 122, 38

\bibitem[van Weeren et al.(2009a)]{vanw09a} van Weeren, R.~J., Intema, H.~T., Oonk, J.~B.~R., R{\"o}ttgering, H.~J.~A., \& Clarke, T.~E.\ 2009, \aap, 508, 1269

\bibitem[van Weeren et al.(2011)]{vanw11} van Weeren, R.~J., R{\"o}ttgering, H.~J.~A., \& Br{\"u}ggen, M.\ 2011, \aap, 527, A114 

\bibitem[van Weeren et al.(2009b)]{vanw09b} van Weeren, R.~J., R{\"o}ttgering, H.~J.~A., Br{\"u}ggen, M., \& Cohen, A.\ 2009, \aap, 508, 75 

\bibitem[Voges et al.(1999)]{RASSBSC} Voges, W., Aschenbach, B., Boller, T., et al.\ 1999, \aap, 349, 389 

\bibitem[Wen et al.(2007)]{wen07} Wen, Z.-L., Yang, Y.-B., Yuan, Q.-R., et al.\ 2007, Chin.\ J.\ Astron.\ Astrophys., 7, 71

\bibitem[Zacharias et al.(2010)]{UCAC3} Zacharias, N., Finch, C., Girard, T., et al.\ 2010, \aj, 139, 2184 

\bibitem[ZuHone et al.(2013)]{zuhone13} ZuHone, J.~A., Markevitch, M., Brunetti, G., \& Giacintucci, S.\ 2013, \apj, 762, 78

\bibitem[ZuHone et al.(2011)]{zuhone} ZuHone, J.~A., Markevitch, M., \& Lee, D.\ 2011, \apj, 743, 16 

\end{thebibliography}
\end{document}